\documentclass[usenatbib,useAMS,letterpaper]{mnras}
\usepackage{float}
\usepackage{graphicx}
\usepackage{fixltx2e}
\usepackage{epstopdf}
\usepackage{ textcomp }
\usepackage{ verbatim}
\usepackage{ctable}
\usepackage{url}
\usepackage{times}
\usepackage{amsmath}
\usepackage{amssymb}
\usepackage{xspace}
\usepackage{mathrsfs} 
\usepackage{tabularx}
\usepackage{fixltx2e}
\usepackage{color}
\usepackage{placeins}
\usepackage{multirow}

\newcommand{\acknowledgments}{\begin{small}\section*{Acknowledgements}\end{small}}

\newcommand\sref[1]{\hyperref[#1]{\S~\ref*{#1}}}
\newcommand\fref[1]{\hyperref[#1]{Fig.~\ref*{#1}}}
\newcommand\Eqref[1]{equation~(\hyperref[#1]{\ref*{#1}})}
\newcommand\tref[1]{\hyperref[#1]{Table~\ref*{#1}}}
\newcommand\aref[1]{\hyperref[#1]{Appendix~A}}

\newcommand{\mr}[1]{\multirow{2}{*}{#1}}

\title[What Fails to Quench?]{The failure of stellar feedback, magnetic fields, conduction, and morphological quenching in maintaining red galaxies}
\author[Su et al.]{
\parbox[t]{\textwidth}{
Kung-Yi Su$^{1}$\thanks{E-mail: ksu@caltech.edu}, Philip F. Hopkins$^{1}$, Christopher C. Hayward$^{2}$,  Xiangcheng Ma$^{1,3}$,  Claude-Andr\'e Faucher-Gigu\`ere$^4$, Du\v san Kere\v s$^5$, Matthew E. Orr$^{1}$, Victor H. Robles$^{6}$
}
\vspace*{6pt} \\
$^1$TAPIR 350-17, California Institute of Technology, 1200 E. California Boulevard, Pasadena, CA 91125, USA\\
$^2$Center for Computational Astrophysics, Flatiron Institute, 162 Fifth Avenue, New York, NY 10010, USA\\
$^3$Department of Astronomy and Theoretical Astrophysics Center, University of California Berkeley, Berkeley, CA 94720, USA\\
$^4$Department of Physics and Astronomy and CIERA, Northwestern University, 2145 Sheridan Road, Evanston, IL 60208, USA\\
$^5$Department of Physics, Center for Astrophysics and Space Sciences, University of California at San Diego, 9500 Gilman Drive, La Jolla, CA 92093, USA\\
$^6$Center for Cosmology, Department of Physics and Astronomy, University of California, Irvine, CA 92697, USA
}

\begin{document}

\long\def\/*#1*/{}
\date{Submitted to MNRAS}

\pagerange{\pageref{firstpage}--\pageref{lastpage}} \pubyear{2017}

\maketitle

\label{firstpage}

\begin{abstract}
The quenching ``maintenance'' and related ``cooling flow'' problems are important in galaxies from Milky Way mass through clusters. We investigate this in halos with masses $\sim 10^{12}-10^{14}\,{\rm M}_{\odot}$, using non-cosmological high-resolution hydrodynamic simulations with the FIRE-2 (Feedback In Realistic Environments) stellar feedback model. We specifically focus on physics present {\em without} AGN, and show that various proposed ``non-AGN'' solution mechanisms in the literature, including Type Ia supernovae, shocked AGB winds, other forms of stellar feedback (e.g.\ cosmic rays), magnetic fields, Spitzer-Braginskii conduction, or ``morphological quenching'' do not halt or substantially reduce cooling flows nor maintain ``quenched'' galaxies in this mass range. We show that stellar feedback (including cosmic rays from SNe) alters the balance of cold/warm gas and the rate at which the cooled gas {\em within} the galaxy turns into stars, but not the net baryonic inflow. If anything, outflowing metals and dense gas promote additional cooling. Conduction is important only in the most massive halos, as expected, but even at $\sim 10^{14}\,{\rm M}_{\odot}$ reduces inflow only by a factor $\sim 2$ (owing to saturation effects and anisotropic suppression). Changing the morphology of the galaxies only slightly alters their Toomre-$Q$ parameter, and has no effect on cooling (as expected), so has essentially no effect on cooling flows or maintaining quenching. This all supports the idea that additional physics, e.g., AGN feedback, must be important in massive galaxies.
\end{abstract}

\begin{keywords}
methods: numerical --- MHD --- galaxy:evolution --- ISM: structure ---  ISM: jets and outflows
\end{keywords}

\section{Introduction} \label{S:intro}

Perhaps the biggest unsolved question in galaxy formation is what explains the ``quenching''\footnote{Throughout this paper, when we refer to ``quenching'' and red galaxies, we exclusively refer to {\em central} galaxies, as opposed to satellite galaxies which can be quenched by a variety of environmental processes (e.g.\ ram pressure and tidal stripping, starvation, strangulation, etc.).} of star formation and maintenance of ``red and dead'' galaxy populations over a large fraction of cosmic time \citep{2003ApJS..149..289B,2003MNRAS.341...54K,2003MNRAS.343..871M,2004ApJ...600..681B,2005ApJ...629..143B,2006MNRAS.368....2D,2009MNRAS.396.2332K,2010A&A...523A..13P,2012MNRAS.424..232W}, at stellar masses $\gtrsim 3-5\times 10^{10}\,{\rm M}_{\odot}$ (above $\sim L_{\ast}$ in the galaxy luminosity function at $z\approx 0$). This is closely related to the classic ``cooling flow problem'': X-ray observations show there exists significant radiative cooling of hot gas in massive ellipticals and clusters with cool-core, indicating cooling times much less than a Hubble time \citep{1994ApJ...436L..63F,2006PhR...427....1P}. Comparing with the inferred cooling flows (reaching up to $\sim 1000\,{\rm M}_\odot\,{\rm yr}^{-1}$ in clusters \citealt{2018ApJ...858...45M}), there are neither sufficient amounts of cold gas (in observed H{\scriptsize I}, e.g.\ \citealt{2011ApJ...731...33M}, or CO, \citealt{2013ApJ...767..153W}), or sufficient star formation rates \citep[SFRs;][]{2001A&A...365L..87T,2008ApJ...681.1035O,2008ApJ...687..899R}, to account for the rapidly-cooling gas (see also \citealt{1976ApJ...208..646S,1977ApJ...215..723C,1978ApJ...224..308M,1991ApJ...376..380C,1994ARA&A..32..277F} for the ``classical'' cooling flow case). Simulations and semi-analytic models which do not suppress these cooling flows, and simply allow the material to cool into galaxies, typically over-produce the observed star formation rates of massive galaxies by at least an order of magnitude \citep[for recent examples, see e.g. the weak/no feedback runs in][]{2007MNRAS.380..877S,2009MNRAS.398...53B,2015MNRAS.449.4105C,2015ApJ...811...73L}.  

To compensate for the observed cooling, there must be some sort of heat source or pressure support. The presence of the shock heated hot-halo can help feedback mechanisms and quench galaxies \citep[e.g.][]{2005MNRAS.363....2K}. However the hot halo itself does not prevent later gas cooling from the cooling flows. The most popular, and perhaps promising, solution is ``feedback'' from an active galactic nucleus (AGN) which can both expel gas from galaxies (shutting down star formation) and inject heat or stirring in the circum-galactic medium (CGM) or intra-cluster medium (ICM), preventing new gas accretion (for recent studies see also e.g \citealt[][]{2012MNRAS.425..605F,2017MNRAS.464.2840A,2017ApJ...847..106L,2018arXiv180506461M,2018arXiv180301444L,2017ApJ...837..149G,2017arXiv171004659W,2017MNRAS.468..751E,2018ApJ...856..115P,2018arXiv180303675Y};  see e.g. \citealt{1998A&A...331L...1S,1999MNRAS.308L..39F,2001ApJ...551..131C,2005ApJ...630..705H,2006ApJS..163....1H,2006MNRAS.365...11C,2007ARA&A..45..117M} for earlier works).  However, despite its plausibility, the detailed physics of AGN feedback (e.g.\ what determines jet energetics and how they transfer energy into the ICM) remains uncertain, as do the relevant ``input parameters'' (e.g.\ kinetic luminosities, duty cycles).

Perhaps as a result, a variety of ``non-AGN'' mechanisms to quench galaxies and keep them red have been proposed in the literature. These generally invoked physics are un-ambiguously present, but play an uncertain role in quenching and the cooling flow problem, including: stellar feedback from shock-heated AGB winds, Type Ia supernovae (SNe), or SNe-injected cosmic rays (CRs); magnetic fields and thermal conduction in the CGM/IGM; or ``morphological quenching'' via altering the galaxy morphology and gravitational stability properties. Our focus in this paper is therefore to attempt a systematic theoretical study of these possibilities, {\em without} considering AGN.

This is important for several reasons: if one (or more) of these mechanisms can, indeed, quench galaxies, this is critical to understand! Even if they do not quench galaxies, they could, in principle, ``help'' by suppressing cooling or star formation (lessening ``requirements'' for AGN). And although many previous studies have claimed AGN feedback is ``necessary'' to explain quenching \citep[see e.g.][in addition to the references above]{1991ApJ...376..380C,2008MNRAS.387...13K,2007ARA&A..45..117M,2008ApJ...681..151C,2015MNRAS.448.1835T}, almost all studies of AGN feedback to date have neglected some or all of these additional processes (often treating e.g.\ stellar feedback in a highly simplified, sub-grid manner). Therefore it is important to understand whether they alter the ``initial conditions'' (e.g.\ typical CGM properties, cooling rates, etc) for AGN feedback. We hope that by studying the ``over-cooling problem'' in global simulations with higher resolution and more detailed physical treatments of the multi-phase ISM and stellar feedback, we can better understand where and how AGN or other feedback, if indeed necessary, must act.

In \sref{S:methods} we summarize the physics considered here, and describe our numerical simulations. Results are presented in \sref{S:results}. We then discuss the effects of each of these physics in turn, in \sref{s:discussion}.

\section{Methodology} \label{S:methods}

Our simulations use {\sc GIZMO} \citep{2015MNRAS.450...53H} \footnote{A public version of this code is available at \href{http://www.tapir.caltech.edu/~phopkins/Site/GIZMO.html}{\textit{http://www.tapir.caltech.edu/$\sim$phopkins/Site/GIZMO.html}}.}, in its meshless finite mass (MFM) mode, which is a Lagrangian mesh-free Godunov method, capturing advantages of grid-based and  smoothed-particle hydrodynamics (SPH) methods. Numerical implementation details and extensive tests are presented in a series of methods papers for e.g.\ the hydrodynamics and self-gravity \citep{2015MNRAS.450...53H}, magnetohydrodynamics \citep[MHD;][]{2016MNRAS.455...51H,2015arXiv150907877H}, anisotropic conduction and viscosity \citep{2016arXiv160207703H,2017MNRAS.471..144S}, and cosmic rays (Chan et al., in prep.). 

Our default simulations use the FIRE-2 implementation of the Feedback In Realistic Environments (FIRE) physical treatments of the ISM and stellar feedback, details of which are given in \citet{2018MNRAS.480..800H,hopkins:sne.methods} along with extensive numerical tests. This follows cooling from $10-10^{10}$K, including the effects of photo-electric and photo-ionization heating by a UV background \cite{2009ApJ...703.1416F} and local source, collisional, Compton, fine structure, recombination, atomic, and molecular cooling. Star formation is allowed only in  gas that is molecular, self-shielding, locally self-gravitating \citep{2013MNRAS.432.2647H}, and above a density $n>100\,{\rm cm^{-3}}$. 
Star particles, once formed, are treated as a single stellar population with metallicity inherited from their parent gas at formation. All feedback rates (SNe and mass-loss rates, spectra, etc.) and strengths are IMF-averaged values calculated from {\small STARBURST99} \citep{1999ApJS..123....3L} with a \citet{2002Sci...295...82K} IMF. The feedback model includes: (1) Radiative feedback including photo-ionization and photo-electric heating, as well as single and multiple-scattering radiation pressure tracked in five bands  (ionizing, FUV, NUV, optical-NIR, IR). (2) Stellar particles continuously lose mass and inject mass, metals, energy, and momentum in the form of OB and AGB winds. (3) Type II and Ia SNe happen stochastically according to the rate mentioned above. Once they occur, the stellar particles lose mass and inject the appropriate mass, metal, momentum and energy to the surrounding gas.

\subsection{Physics Surveyed}
\label{S:physics}

\subsubsection{Stellar Feedback: Young/Massive Stars}
\label{S:feedback}

Feedback from massive stars is un-ambiguously crucial to galaxy evolution. In the last decade, with progress in modeling stellar feedback, simulations of $\lesssim L*$ galaxies \citep[see e.g.][]{2007MNRAS.374.1479G,2009ApJ...695..292C,2012MNRAS.423.2374U,2011MNRAS.417..950H, 2012MNRAS.421.3488H,2012MNRAS.421.3522H,2015MNRAS.454.2691M,2015arXiv150900853A,2015arXiv151005644H,2014ApJ...786...64K} are now able to produce reasonably realistic galaxy populations, without the runaway collapse and star formation that occurs absent feedback. However in these (mostly star-forming) lower-mass galaxies, feedback is dominated by young, massive stars (e.g.\ radiation and OB winds from massive stars, Type-II SNe). Given the {\em observed} low specific star formation rates (SSFRs) in quenched systems \citep[e.g.\ $\lesssim 10^{-11}\,{\rm yr^{-1}}$ for $10^{14}{\rm M}_\odot$ halos;][]{2006MNRAS.366....2W,2013MNRAS.428.3306W,2014arXiv1401.7329K}, the number of massive stars is very low, so it seems unlikely this can maintain a quenched galaxy without (paradoxically) a much larger SFR. But these physics must be present whenever star formation does occur, so we include them with the methods described above.

\subsubsection{Stellar Feedback: SNe Ia}

At the observed low SSFRs of massive (quenched) galaxies, the SNe Ia rate (including both prompt and delayed populations) is much larger than the core-collapse rate, giving a time-averaged energy-injection rate $\sim 10^{41.5}\,{\rm erg\,s^{-1}}\,(M_{\ast}/10^{11}\,_{\odot})$, which can be comparable to the cooling rates in some systems. Since these come from old populations, and are distributed smoothly in space and time, it has been proposed that they could be an important CGM/ICM heating mechanism \citep[e.g.][and references therein]{2005ApJ...628..205T,2009MNRAS.398.1468T,2012MNRAS.420.3174S}. We include Ia's following the FIRE-2 method described above, using with the rates from \citet{2006MNRAS.370..773M} (including both the prompt and delayed components), assuming $10^{51}\,{\rm erg}$ per event. Note that although there has been considerable debate about Ia rates, it has focused on the prompt component, which is unimportant for our conclusions.

\subsubsection{Stellar Feedback: AGB Winds}

AGB winds from old stellar populations return a significant fraction of the stellar mass, but have  low launch velocities $\sim 10\,{\rm km\,s^{-1}}$ and correspondingly negligible kinetic luminosities. However, \cite{2015ApJ...803...77C} note that if the AGB stars are moving through the ambient gas medium with large velocity dispersions $\gtrsim 300\,{\rm km\,s^{-1}}$, the kinetic luminosities and post-shock temperatures are greatly elevated in the wind bow shocks, and this can suppress cooling and inject energy well above the Ia rate. Crucially, our default FIRE-2 models account in detail for the relative star-gas velocity when injecting stellar mass loss of any kind (AGB or OB winds or SNe), in an exactly conservative manner, as described and tested in \citet{hopkins:sne.methods}.

\subsubsection{Magnetic Fields, Conduction \&\ Viscosity}
\label{S:microphysics}

Magnetic fields can, in principle, directly suppress cooling flows via providing additional pressure support \citep{1990ApJ...348...73S,1996ARA&A..34..155B,2012MNRAS.422.2152B}, although they have limited effects on global star formation properties of sub-$L_{\ast}$ galaxies \citep{2017MNRAS.471..144S}. They can also non-linearly influence essentially all the gas dynamics. 

Thermal conduction can carry heat from the outer parts of hot halos into cool inner cores, and so might serve as an important heating mechanism
\citep{1981ApJ...247..464B,1983ApJ...267..547T,2002MNRAS.335L...7V,2002MNRAS.335L..71F,2003ApJ...582..162Z,2003ApJ...596..889K,2004MNRAS.347.1130V,2004ApJ...606L..97D,2006MNRAS.367.1121P,2011ApJ...740...28V,2014MNRAS.439.2822W}. However under the conditions in the CGM/ICM, this cannot be considered in the absence of MHD, as the conduction is highly anisotropic. Convective instabilities driven by anisotropic (Spitzer-Braginskii) conduction \citep{1953PhRv...89..977S,1988xrec.book.....S,2003ApJ...582..162Z,2015ApJ...798...90Z,2016MNRAS.458..410K} along magnetic field lines, including the heat-flux-driven buoyancy instability \citep[HBI;][]{2008ApJ...673..758Q,2008ApJ...677L...9P} and the magnetothermal instability \citep[MTI;][]{2000ApJ...534..420B,2008ApJ...688..905P}, may further change the magnetic configuration and conduction time scale \citep{2009ApJ...703...96P} or even drive turbulence and provide extra pressure support or mixing \citep{2012MNRAS.419L..29P}. It is also been argued that conduction can help AGN feedback quench galaxies more effectively \citep[e.g.][]{2017ApJ...837L..18K}.

We therefore consider a set of additional fluid ``microphysics'' runs, with ideal MHD and physical (temperature-dependent, fully-anisotropic) Spitzer-Braginskii conduction and viscosity (we assume the perpendicular transport coefficients are vanishingly small).  The implementation is identical to \citet{2017MNRAS.471..144S}.

\subsubsection{Cosmic Rays (not from AGN)}

Cosmic rays (CRs) can provide additional pressure support to gas, drive galactic outflows, and heat the CGM/ICM directly via hadronic and streaming losses \citep{2008MNRAS.384..251G,2010ApJ...720..652S,2011A&A...527A..99E,2011ApJ...738..182F,2013MNRAS.434.2209W,2013MNRAS.432.1434F,2017ApJ...834..208R,2017ApJ...844...13R,2013ApJ...779...10P,2017MNRAS.465.4500P,2017MNRAS.467.1449J,2017MNRAS.467.1478J,2018MNRAS.475..570J}. As a result several of the studies above suggest they can help quench star formation; however this is usually in the context of CRs from AGN. Here we wish to explore non-AGN mechanisms, so we consider simulations adopting the CR physics and numerical implementation described in (Chan et al., in prep.). This CR treatment includes including streaming (at the local Alfv\'en speed or sound speed, whichever is larger, $v_{\rm st}\sim\sqrt{v_{\rm Alf}^2+v_{\rm c}^2}$, with the appropriate streaming loss term, which thermalizes, following \citealt{Uhlig2012}), diffusion (with a fixed diffusivity $\kappa_{\rm cr}$), adiabatic energy exchange with the gas and cosmic ray pressure in the gas equation of motion, and hadronic and Coulomb losses (following \citealt{2008MNRAS.384..251G}). We follow a single energy bin (i.e.\ GeV CRs, which dominate the pressure), treated in the ultra-relativistic limit. Streaming and diffusion are fully-anisotropic along field lines. CRs are injected in SNe (a fixed fraction $\epsilon_{\rm cr}=0.1$ of each SNe energy goes into CRs; see e.g. \citealt{2017MNRAS.465.4500P,2017ApJ...847L..13P}). In Chan et al. (in prep.), we show that matching observed $\gamma$-ray luminosities, in simulations with the physics above requires $\kappa_{\rm cr}\sim 10^{29}\,{\rm cm^{2}\,s^{-1}}$ (in good agreement with detailed CR transport models that include an extended gaseous halo around the Galaxy,  \citep[see e.g.][]{1998ApJ...509..212S,2010ApJ...722L..58S,2011ApJ...729..106T},
so we adopt this as our fiducial value, but discuss variations below. 

We note that in addition to SNe shocks, the other major non-AGN source of CRs of interest here is shocks from cosmological large scale structure (LSS) formation/accretion. Since our simulations are not fully-cosmological, this is not directly accounted for. \footnote{We implicitly effectively model this in the CR energy density in our initial conditions by assuming equipartition with magnetic energy}

\subsubsection{Morphological Quenching}

Finally, \cite{2009ApJ...707..250M} and \cite{2009ApJ...703..785D} described a scenario they referred to as ``morphological quenching,'' whereby quenching could be accomplished (SF suppressed) simply by altering a galaxy's morphology. Specifically they argued that turning a stellar disk into a more gravitationally stable spheroid would raise the Toomre-$Q$ and stabilize the gas against fragmentation/star formation. This involves no new physics beyond those above (our simulations easily resolve $Q$ and the vertical scale-heights and gravitational fragmentation of the cold gaseous disks), but rather different galaxy initial conditions given the same halo properties.

\subsection{Initial conditions} 
\label{s:ic}

It is important to note that the over-cooling problem exists over several orders of magnitude in halo mass, not just at $\sim L^{\ast}$ where most galaxies first quench, or in massive clusters where the classical ``cooling flow problem'' is defined. We therefore consider three fiducial initial conditions (ICs), with halo masses of $1.5\times10^{12}$ (m12), $10^{13} $ (m13) and $8.5\times10^{13} {\rm M}_\odot$ (m14), respectively. The DM halo, stellar bulge, stellar disc are set  following \cite{1999MNRAS.307..162S}.
We assume a spherical, isotropic, NFW \citep{1996ApJ...462..563N} profile dark matter halo (scale lengths (20.4, 93, 218.5) kpc), and  \cite{1990ApJ...356..359H} profile stellar bulge (scale lengths (1, 2.8, 3.9) kpc). 
We also assume exponential, rotation-supported gas and stellar disks (scale lengths (6, 2.8, 3.9) kpc and (3, 2.8,  3.9) kpc, respectively; scale-height (0.3, 1, 1.4) kpc for both, gas temperatures initialized to pressure equilibrium \citealt{2000MNRAS.312..859S}), and an extended spherical, hydrostatic gas halo with a beta profile (scale-radius (20.4, 9.3, 21.85) kpc and $\beta=(0.5,0.43, 0.5)$). \footnote{The hot halo gas rotates at a fix fraction of the circular velocity, which is twice the DM halo streaming velocity in \cite{1999MNRAS.307..162S}. It is then $\sim10-15\%$  rotation-supported, and $\sim 85-90\%$ thermal-pressure-supported.} All the initial conditions except the m13 case are run adiabatically (no feedback, no cooling) for at least 50 Myr before putting into use, to ensure stability. \footnote{This is necessary for the runs with radially-dependent super-Lagrangian refinement scheme (m14), but less relevant for the others.}
The properties are summarized in \tref{tab:ic}. In the m12 cases, we explicitly test the effect of different stellar morphology on the cooling, so besides the fiducial (disc-dominated) m12 IC, we also construct bulge-dominated and gas-disk-free ICs (m12 Bulge and m12 Bulge-NoGD).

The initial conditions are set up to be similar to typical cooling-core systems observed, insofar as this is possible. For example, m12 is a Milky Way-mass galaxy, where the hot gas halo, roughly follows the observed Milky Way profile estimated in \cite{2013ApJ...770..118M,2015ApJ...800...14M} and \cite{2017ApJ...836..243G}, except that we assume a universal baryonic fraction (0.16) inside twice the virial radius (we do not allow for missing baryons as suggested in the aforementioned papers). It is possible that the solution to over-cooling involves expulsion of a large fraction of the Universal baryonic mass; however our intention here is to see if this does occur, not to put it in ``by hand'' (moreover, direct observations indicate the full baryon content does exist within similar radii, in external systems, see \citealt{2013A&A...557A..52P,2015ApJ...808..151G,2017arXiv171208619L}). 
Our m13 is the elliptical galaxy (Ell) in \cite{2017MNRAS.471..144S}\footnote{There is minor typographical error in the tabulated Ell properties in \cite{2017MNRAS.471..144S}. The values in \tref{tab:ic} are correct.} The mass and radial distribution of  gas, stars and dark mater are consistent with the observations of similar-mass  halos compiled in \citet{2012ApJ...755..166H} and \citet{2016MNRAS.455..227A} . Our 
 m14 is initialized as as cool core cluster, with a massive central elliptical galaxy, by design. The halo properties and  profiles are consistent with typical observed systems of the same mass \citep{2012ApJ...748...11H,2013MNRAS.436.2879H,2013ApJ...775...89S,2015ApJ...805..104S}

\begin{figure}
\centering
\includegraphics[width=8.5cm]{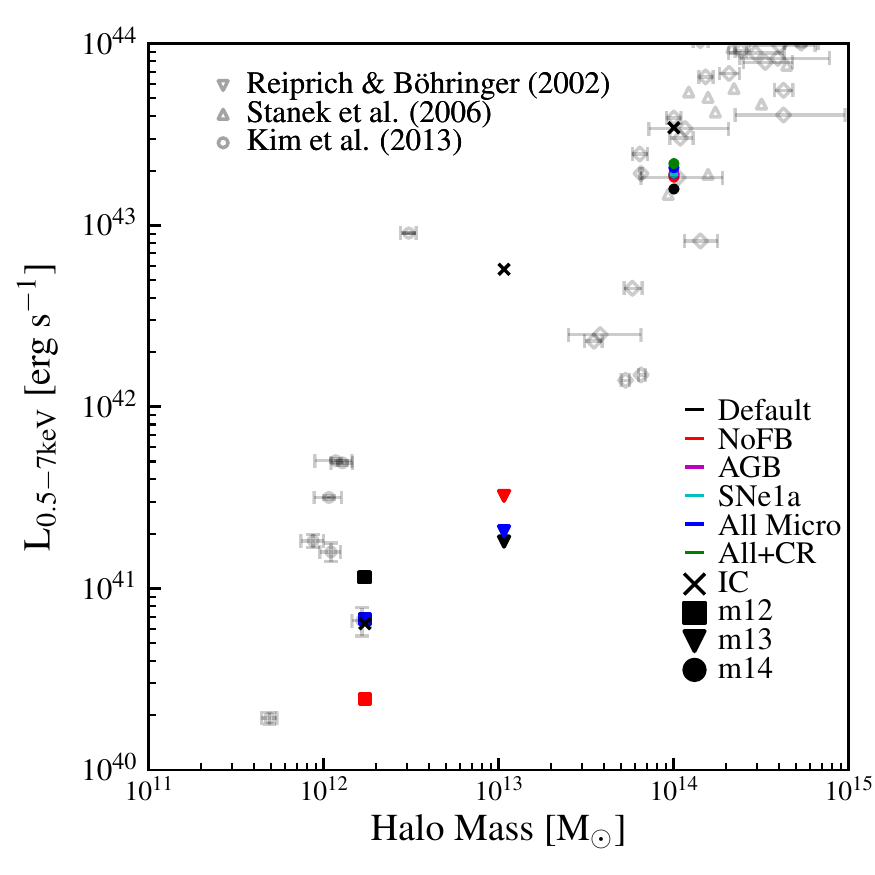}
\caption{The X-ray luminosity (0.5-7keV) of our initial conditions and the average luminosity of the last 100 Myr of each run are plotted on the  X-ray luminosity - halo mass plane in comparison to the observations.  All our runs lie reasonably within the scatter of the observed  X-ray luminosity - halo mass  relation.  In m12, the low halo temperature means the X-ray luminosity is significantly influenced by SNe heating; in m13 \&\ m14, the effects of stellar FB on the X-ray luminosity are small (most comes from the initial hot halo). Magnetic fields and CR feedback have little effect on the X-ray luminosity.}
\label{fig:xray}
\end{figure}

The comparison of the X-ray luminosity of our halo to the observations is plotted in the  X-ray luminosity - halo mass plane in \fref{fig:xray}.\footnote{The X-ray luminosity in our simulation is measured over $0.5-7$\,keV. That from \citet{2002ApJ...567..716R} and \citet{2002ApJ...567..716R} is from $0.5-2.4$\,keV, while that from \citet{2013ApJ...776..116K} is measured from $0.3-8$\,keV, but these are corrected given their median estimated spectral slope to the same range we measure. Given that most of the X-ray emission halo in this mass range in below $2\,$keV, and we are not trying to reproduce a specific halo, the comparison here is not particularly sensitive to this.} The luminosity  is calculated  using the same method in \cite{2018MNRAS.478.3544R}, in which the cooling curve is calculated  for the photospheric solar abundances \citep{2003ApJ...591.1220L}, using the spectral analysis code SPEX \citep{1996uxsa.conf..411K} in the same way of \cite{2009A&A...508..751S} and scaled according to the local  hydrogen, helium, and metal mass fractions.  All our initial conditions have cooling luminosity within the scatter of the observed X-ray luminosity - halo mass relation \citep{2002ApJ...567..716R,2006ApJ...648..956S,2006MNRAS.366..624B,2013ApJ...776..116K,2015MNRAS.449.3806A}.

In all runs, unless otherwise noted, the initial metallicity is set to solar ($Z=0.02$) at the core, and drops to $Z=0.001$  at larger radii. \footnote{The metallicity scales as $Z_\odot(0.05+0.95/(1+(r/R_c)^{1.5}))$, where $R_c$ is set to (20,10,20) kpc for (m12, m13, m14).} The m14 `Low Metal' run is set to $Z=0.001$ uniformly. For runs with magnetic fields, the initial magnetic field strength of the gas in the core is set to $0.03\,\mu {\rm G}$, $0.1\,\mu {\rm G}$ and $0.3\,\mu {\rm G}$ for m12, m13 and m14, respectively (roughly according to $m_{\rm vir}^{1/2}$).  The initial field configuration is azimuthal, and decays as a function of radius.\footnote{Magnetic field strength scales as $(1/(1+(r/R_c)^{2}))^\beta$, where $R_c=(20, 10, 20)$ kpc, and $\beta=(0.375, 1, 0.375)$ for (m12, m13, m14)} For runs with cosmic rays, the initial CR energy density is set to be in local equipartition with the initial magnetic energy density, at all positions.

Given that m14 is very massive (with much of the gas mass in the extended, non-cooling halo at radii $\sim $Mpc),  resolving it with a uniform gas mass resolution is computationally formidable and not necessary for the convergence of FIRE stellar feedback. We therefore adopt a radially-dependent super-Lagrangian refinement scheme in this case: the target gas mass resolution is set to $8000\,{\rm M}_{\odot}$ inside $r < 10\,$kpc, and increases smoothly $\propto r$ outside  this radius up to a maximum $2\times 10^{6}\,{\rm M}_{\odot}$ at $\sim 300\,$kpc. Gas resolution elements are automatically merged or split appropriately if they move inward/outward, to maintain this mass resolution (to within a factor $=2$ tolerance) at all times.

Most simulations have been rerun with different resolutions, with the initial  mass resolution differing by at least 2 orders of magnitude. The conclusions are robust in this resolution range, with resolution studies provided in \aref{appendix}. The list of runs are summarized in \tref{tab:run}. We note that the m14 runs with the ``extended fluid microphysics'' set or cosmic rays are more expensive, and are therefore run with lower resolution, but within the range where our results appear robust (see \aref{appendix}).

\begin{table*}
\begin{center}
 \caption{Simulation properties}
 \label{tab:ic}
 \begin{tabular*}{\textwidth}{@{\extracolsep{\fill}}lccccccccccccccc}
 \hline
\hline
&\multicolumn{2}{c}{\underline{Resolution}}&\multicolumn{3}{c}{\underline{DM halo}}&&\multicolumn{2}{c}{\underline{Stellar Bulge}}&\multicolumn{2}{c}{\underline{Stellar Disc}}&\multicolumn{2}{c}{\underline{Gas Disc}}&\multicolumn{2}{c}{\underline{Gas Halo}}  \\
$\,\,\,\,$Model  &$\epsilon_g$ &$m_g$        &$M_{\rm halo}$   &$c$            &$V_{\rm Max}$    &$M_{\rm bar}$    &$M_b$ 
                 &$a$          &$M_d$        & $r_d$             &$M_{gd}$       &$r_{gd}$         &$M_{gh}$         &$r_{gh}/r_{dh}$     \\
                 &(pc)         &(M$_\odot$)  &(M$_\odot$)      &               &(km/s)           &(M$_\odot$)      &(M$_\odot$) 
                  &(kpc)        &(M$_\odot$)  &(kpc)            &(M$_\odot$)    &(kpc)            &(M$_\odot$)      &\\
\hline
$\,\,\,\,$m12        &1       &8e3           &1.5e12           &12             &174              &2.2e11           &1.5e10   
                     &1       &5e10          &3                &5e9            &6                &1.5e11           &1        \\
$\,\,\,\,$m12 Bulge  &1       &8e3           &1.5e12           &12             &174              &2.2e11           &6e10   
                     &1.6     &5e9           &1.4              &5e9            &6                &1.5e11           &1        \\                     
$\,\,\,\,$m12 Bulge-NoGD&1   &8e3           &1.5e12           &12             &174              &2.2e11           &6e10   
                     &1.6     &5e9           &1.4              &5e8            &2.8              &1.6e11           &1        \\
$\,\,\,\,$m13        &3       &5e4           &1.0e13           &6              &240              &7.2e11           &1e11   
                     &2.8     &1.4e10        &2.8              &5e9           &2.8              &6e11              &0.1             \\
$\,\,\,\,$m14 MR     &1       &3e4*           &8.5e13           &5.5            &600              &1.52e13          &2e11
                     &3.9     &2e10          &3.9              &1e10           &3.9              &1.5e13           &0.1             \\          
$\,\,\,\,$m14 HR     &1       &8e3*           &8.5e13           &5.5            &600              &1.52e13          &2e11
                     &3.9     &2e10          &3.9              &1e10           &3.9              &1.5e13           &0.1             \\   
\hline 
\hline
\end{tabular*}
\end{center}
\begin{flushleft}
Parameters of the galaxy models studied here (\sref{s:ic}):\\
(1) Model name. The number following `m' labels the approximate logarithmic halo mass. m12 is a disc dominant halo, while m12 Bulge, m13 and m14 are bulge-dominant. The run labeled NoGD have an order of magnitude smaller gas disc.
(2) $\epsilon_g$: Gravitational force softening for gas (the softening for gas in all simulations is adaptive, and matched to the hydrodynamic resolution; here, we quote the minimum Plummer equivalent softening).
(3) $m_g$: Gas mass (resolution element). There is a resolution gradient for m14, so its $m_g$ (with *) is the mass of the highest resolution elements.
(4) $M_{\rm halo}$: Halo mass. 
(5) $c$: NFW halo concentration.
(6) $V_{\rm max}$: Halo maximum circular velocity.
(7) $M_{\rm bar}$: Total baryonic mass. It is the sum of gas, disc, bulge and stellar mass for isolated galaxy runs, and the sum of gas and stellar mass in the cosmological runs within 0.1 virial radius.
(8) $M_b$: Bulge mass.
(9) $a$: Bulge scale-length (Hernquist profile).
(10) $M_d$ : Stellar disc mass. For CosmoMW and CosmoDwarf runs, this is the total stellar mass within 0.1 virial radius.
(11) $r_d$ : Stellar disc scale length (exponential disc).
(12) $M_{gd}$: Gas disc mass. 
(13) $r_{gd}$: Gas disc scale length (exponential disc).
(14) $M_{gh}$: Gas halo mass. 
(15) $r_{gh}/r_{dh}$: Gas halo scale length (beta profile) over dark matter scale length.
\end{flushleft}
\end{table*}

\begin{table}
\begin{center}
 \caption{List of runs}
 \label{tab:run}
 \begin{tabular}{ccccc}
 \hline
\hline
Model     &Feedback   & Microphysics    &CR     \\
\hline
\underline{\bf m12}\\
NoFB       &None          &Hydro                &no       \\
Default    &FIRE 2      &Hydro                &no            \\
All Micro  &FIRE 2      &MHD+Viscosity+Conduction               &no        \\
Default-Bulge        &FIRE 2      &Hydro               &no        \\
NoFB-Bulge      &\mr{FIRE 2}      &\mr{Hydro}               &\mr{no}        \\
-NoGD\\
\hline 
\underline{\bf m13}\\
NoFB       &None          &Hydro                &no       \\
Default    &FIRE 2      &Hydro                &no            \\
All Micro  &FIRE 2      &MHD+Viscosity+Conduction               &no        \\
\hline 
\underline{\bf m14 HR}\\
NoFB       &None          &Hydro                &no       \\
Default    &FIRE 2      &Hydro                &no            \\
AGB       &AGB winds only   &Hydro                &no       \\
SNeIa     &Type Ia SNe only &Hydro                &no       \\
Low Metal   &FIRE 2      &Hydro                &no\\
\hline 
\underline{\bf m14 MR}\\
Default    &FIRE 2      &Hydro                &no            \\
All Micro  &FIRE 2      &MHD+Viscosity+Conduction               &no        \\
All+CR     &FIRE 2      &MHD+Viscosity+Conduction               &yes       \\
\hline 
\hline
\end{tabular}
\end{center}
\end{table}

\section{Results} \label{S:results}

\subsection{Gas Masses \&\ Phases in Cores} \label{S:phase}
\begin{figure*}
\centering
\vspace{0.15cm}
\includegraphics[width=16cm]{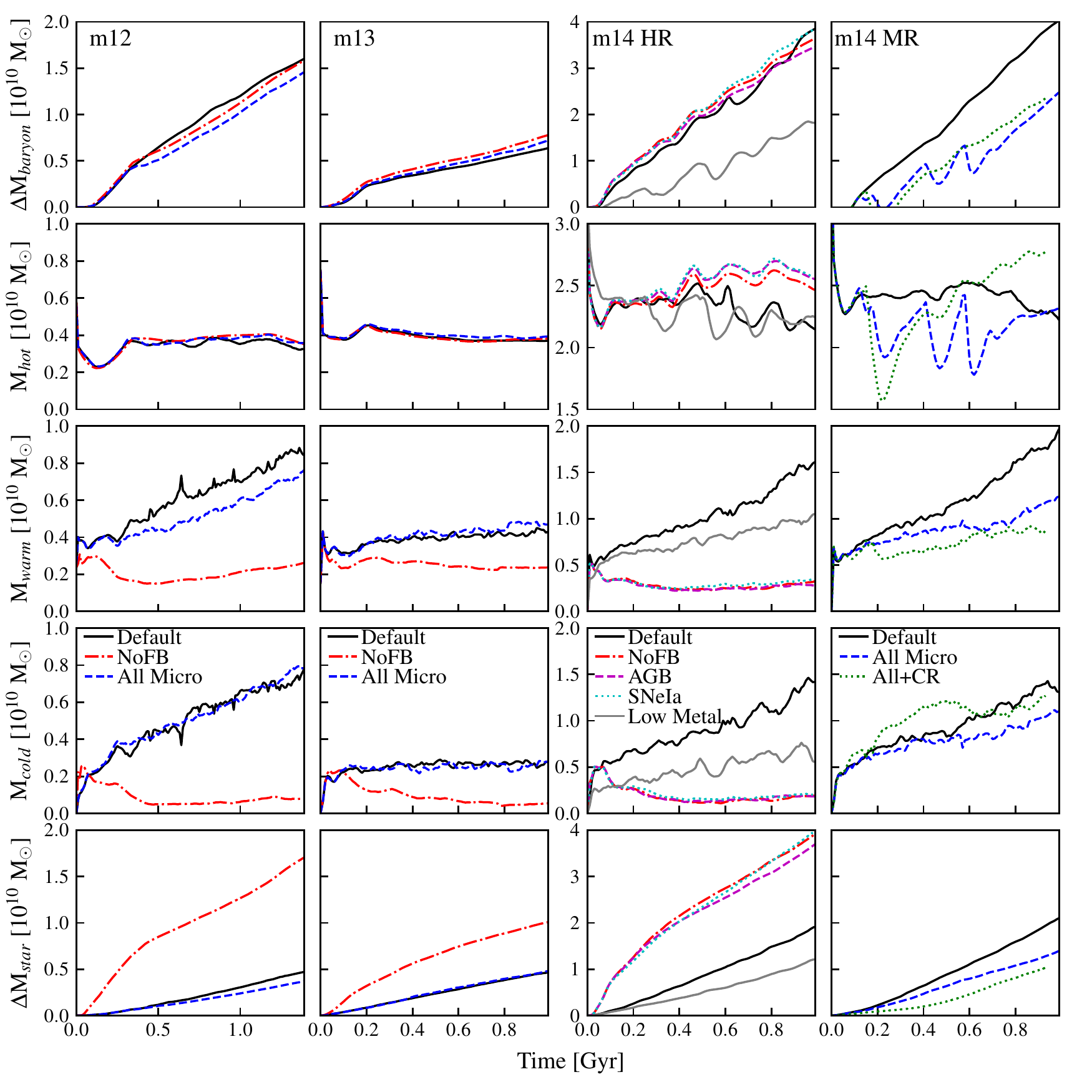}
\vspace{0.15cm}
\caption{Cooling flows in different phases as a function of time: {\bf Top row:} The baryonic mass variation within 30 kpc ($\Delta M_{\rm baryon}$). {\bf 2nd row:} The total hot gas ($>10^6$ K) mass within 30 kpc  ($ M_{\rm hot}$). {\bf 3rd row:} The total warm gas ($8000-10^6$ K) mass within 30 kpc  ($ M_{\rm warm}$). {\bf 4th row:} The total cold gas ($<8000$ K) mass within 30 kpc  ($ M_{\rm cold}$). {\bf Bottom row:} The variation of stellar mass  within 30 kpc  ($\Delta M_{\rm star}$).
In the `Default' run, the cold and warm gas mass within 30kpc grows rapidly as gas cools, but in the `NoFB' runs, any cooled gas almost immediately forms star, so only stellar mass increases. 
Both the`AGB' and `SNeIa' runs  behave roughly similarly to the `NoFB' runs, indicating most of the FB comes from massive stars in these runaway-cooling simulations.
Without metal-line cooling (`Low Metal' m14 run), the built-up cold and warm gas are all suppressed by a factor of 1.5-2. Both the `All Micro' and `All+CR' runs are similar to the `Default' run in lower mass cases, but have a modest difference in the m14 halo, where conduction suppresses cooling flow by a factor of 2, and CR feedback suppresses SF and allows the build-up of additional cold gas.}
\label{fig:hotgas}
\end{figure*}

\fref{fig:hotgas} plots the baryonic mass within $<30\,$kpc ($M^{30\,{\rm kpc}}_{\rm baryon}$). As gas cools, this increases, with rate $\dot{M}^{30\,{\rm kpc}}_{\rm baryon} / {\rm M}_{\odot}\,{\rm yr^{-1}}\sim (12,\,6,\,40)$ in (m12,m13,m14): there is a competition at increasing mass between higher temperatures (lower cooling rates per particle) and simply larger gas masses available to cool (so this decreases slightly from m12 to m13, then rises rapidly to m14). This is also partially because the viral temperature of `m13' is roughly at the minimum of the cooling curve (a few times $10^6$ K).  Gas with $T>10^{6}$\,K is mostly hot halo gas from the ICs and is replenished (from larger radius) at small radii as it cools (only a small fraction comes from stellar feedback) so the ``hot'' gas mass evolves only weakly. In our default runs the cold and warm gas mass inside $<30\,$kpc grows rapidly as gas cools. In the ``NoFB'' runs this does not appear only because that cold/warm gas turns into stars almost immediately (in $\sim 1$ local free-fall time); stellar FB slows the cold gas consumption time to $\sim1-2$\,Gyr.

The ``All Micro'' and ``All+CR'' runs are similar to our Default at lower masses, and produce modest effects at m14-mass, with conduction lowering cool inflow rates by a factor $\sim2$, and CRs suppressing SF in cool gas (and building up additional cold gas) by a similar factor at early times. ``AGB'' and ``SNeIa'' (runs using {\em only} these stellar FB mechanisms) are similar to NoFB, indicating most of the FB comes from massive stars in these runaway-cooling simulations. Our ``Low Metal'' run suppresses the buildup of cold+warm gas by a factor $\sim 1.5-2$, owing to the lack of metal-line cooling from the hot gas.

\vspace{-0.5cm}
\subsection{Star Formation Rates} \label{S:sfr}
\begin{figure*}
\centering
\includegraphics[width=17cm]{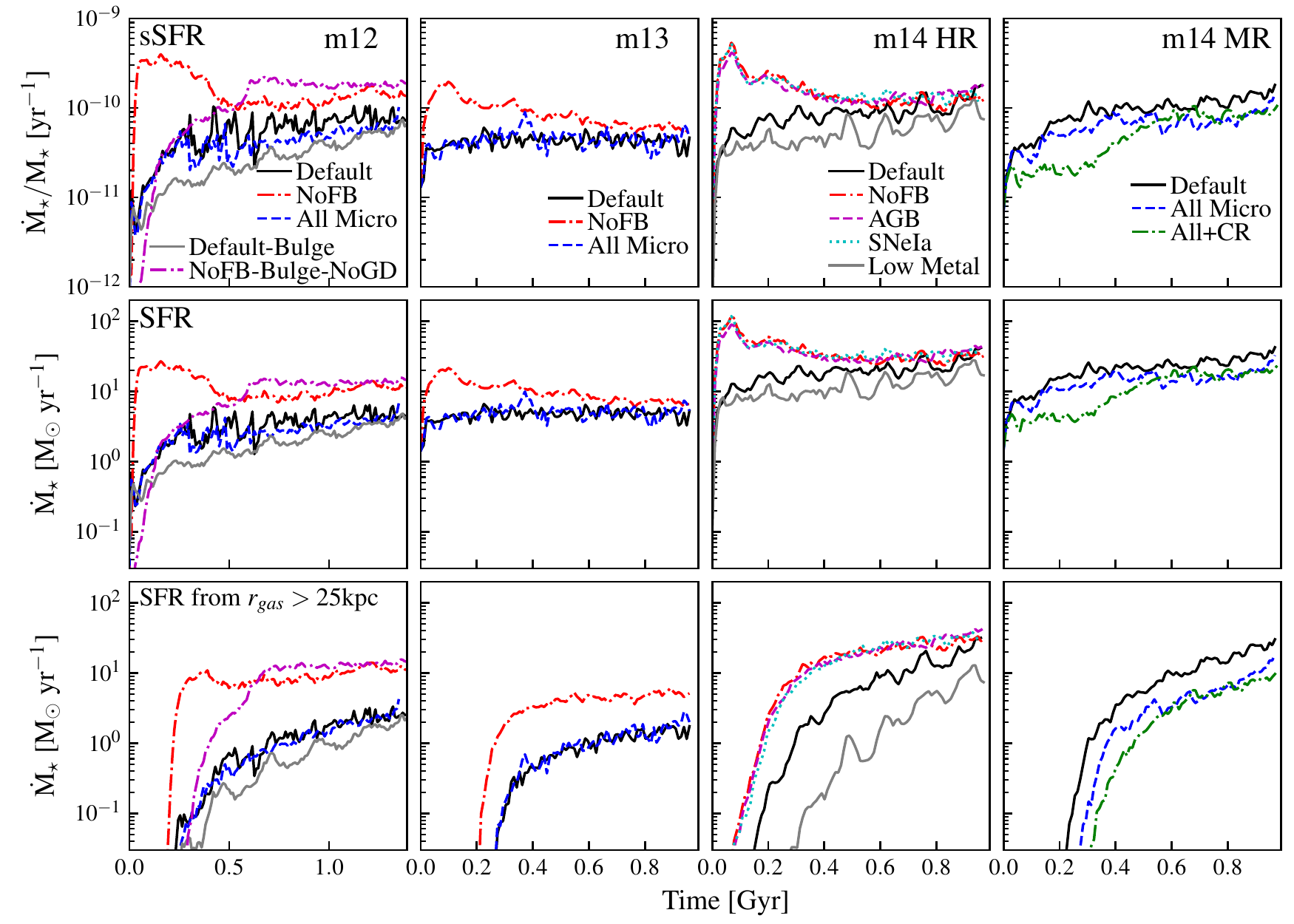}
\caption{{\bf Top row: } Specific star formation rate (sSFR) as a function of time. {\bf Second row:} Star formation rate (SFR) as a function of time, averaged over 100 Myr. {\bf Third row: } Star formation rates from gas initially at radii larger than 25 kpc (fueled by cooling flow). 
None of the galaxies are quenched.  The `NoFB' runs (similarly also the `AGB' and `SNe1a' runs) have an initial rapid star burst, where the initial core gas cools rapidly and forms stars. After $\sim 100$ Myr, gas depletion in the core lowers the initial core SF, and then the subsequent SF tracks the cooling flow gas. The other runs (`Default', `All Micro', `All+CR') which initially form fewer stars, and preserve larger gas reservoirs as gas cools, but they all eventually have cooling-regulated SFRs resembling the SFR of the `NoFB' runs. Suppressing metal-line cooling  (`Low Metal' run) lowers the SFR by a factor of 2. With or without feedback, the saturated SFRs of the disc-dominant m12 runs (`Default' and `NoFB') are very similar to the corresponding bulge-dominant runs (`Default-Bulge' and `NoFB-Bulge-NoGD'), indicating that the morphology of the stellar distribution has little effect on star formation.}
\label{fig:sfr}
\end{figure*}

\fref{fig:sfr} shows SFRs and sSFRs (averaged in rolling $10\,$Myr bins): typical $\dot{M}_{\ast} \sim 2-5 \,{\rm M}_{\odot}\,{\rm yr^{-1}}$ in m12 and m13, and $\sim 20-40 \,{\rm M}_{\odot}\,{\rm yr^{-1}}$ in m14. In sSFR m12 \&\ m14 have $\dot{M}_{\ast}/M_{\ast} \sim 10^{-10}\,{\rm yr^{-1}}$, m13 $\sim 3-5\times10^{-11}\,{\rm yr^{-1}}$: none of is ``quenched.'' 
In fact, in m14, the SFRs and cooling flows are accelerating, indicating development of a stronger cooling flow with time.
The ``NoFB'' runs have an early, rapid burst, where gas in the initial core undergoes runaway collapse and SF, until $\sim 100\,$Myr when gas depletion in the core lowers the SFR and subsequent SF comes from gas initially at larger radii, tracking the cooling rate (again, the only-``AGB'' and only-``SNeIa'' runs resemble NoFB). The Default, ``All Micro,'' and ``All+CR'' runs initially turn less gas into stars, but this leads to their preserving a larger gas reservoir as cooling continues, until eventually the SFRs are similar to ``NoFB'' (cooling-regulated). Again, effectively removing metal-line cooling in the ``Low Metal'' run reduces cooling and late SFRs by a factor $\sim2$.

Note m13 \&\ m14 are entirely bulge-dominated, but still feature high sSFRs. In m12 we explicitly test different initial stellar morphologies: the ``Default-Bulge'' (bulge-dominated) and ``Default'' (disk-dominated) runs give similar SFRs (with or without stellar FB). We also compare ``NoFB-Bulge-NoGD,'' a run with no feedback and no {\em gas} disk initially, so the only SF can come from gas cooling from large radius. Even in this case, while the initial SFR is lower owing to the lack of initial gas supply, the SFR saturates to the same value as the ``NoFB'' (disk-dominated, with gas disk) run.

\vspace{-0.5cm}
\subsection{Cooling Times \&\ Thermal Stability} \label{S:cool_t}

\begin{figure*}
\centering
\includegraphics[width=17cm]{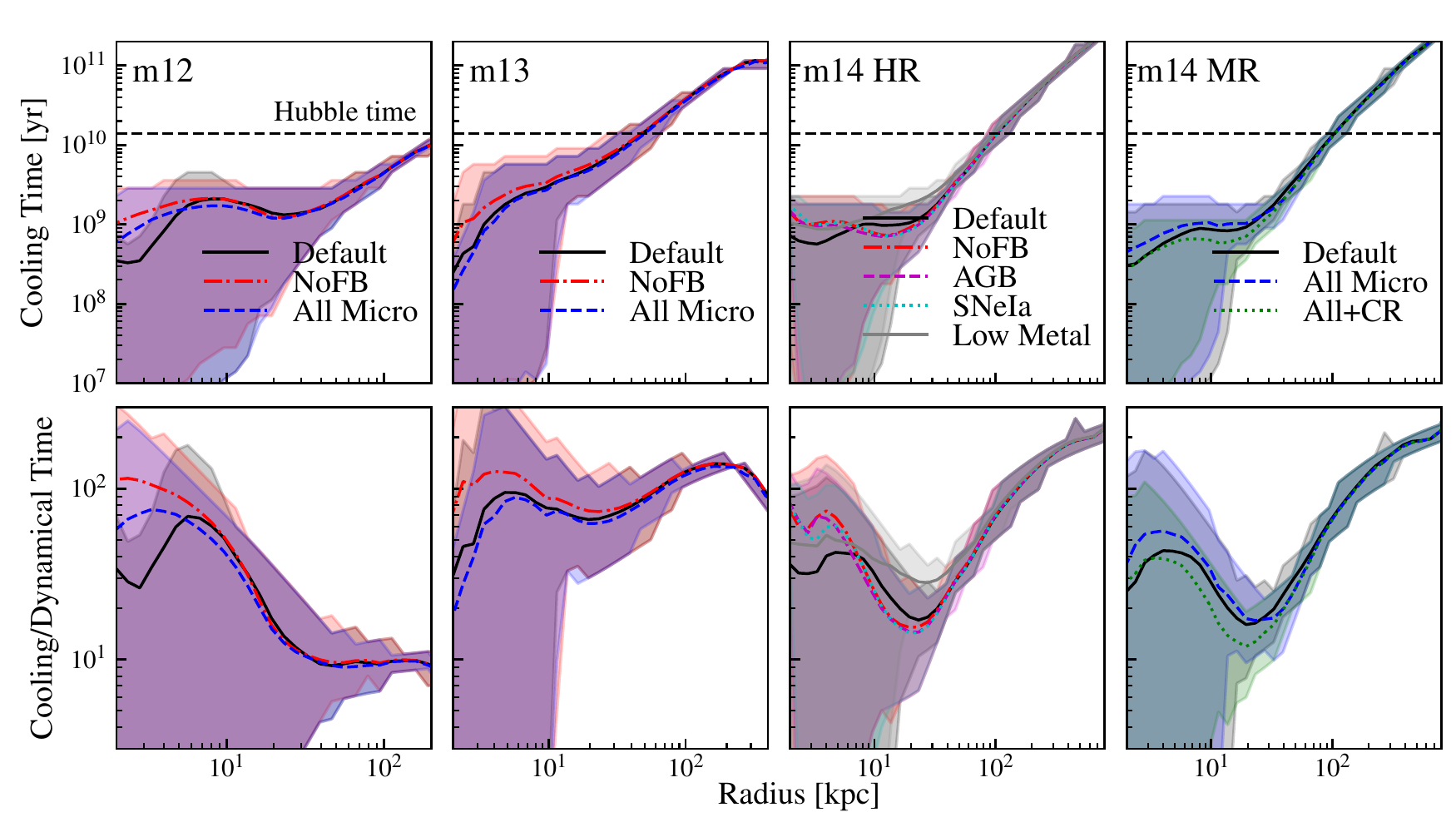}
\caption{{\bf Top row:} Cooling time of  gas hotter than $10^5 {\rm K}$,  as a function of radius. Black dashed lines indicate the Hubble time. {\bf Bottom row:} Ratio of cooling time to dynamical time, as a function of radius. The lines are mass-weighted averages and the shaded region covers 10-90\%  of the distribution. All the quantities are averaged over the 80-90th Myr since the beginning of the simulations. 
The `Default' runs have lower $\tau_{c}/\tau_{d}$ than the `NoFB' runs within $\sim 10 $kpc since FB enriches the gas with metals. Both the `AGB-only' and `SNe1a-only' runs resemble the `NoFB' run. Conduction and  CR feedback do not have a significant effect. Suppressing metal-line cooling (`Low Metal' run) increases $\tau_{c}$ by a factor of 2.}
\label{fig:coolingtime}
\end{figure*}

\fref{fig:coolingtime} shows the cooling time ($\tau_{c}$) of gas hotter than $10^{5}$\,K as a function of radius averaged over the 80-90th Myr since the beginning of the simulations. Within $\sim (200,\,40,\,100)\,$kpc in (m12,m13,m14) cooling times are short compared to the Hubble time (at large radii temperatures are higher, metallicities and densities lower, so $\tau_{c}$ rise rapidly). The ratio of cooling time to dynamical time ($\tau_{c}/\tau_{d}$) is also shown\footnote{$\tau_{c}=E_{\rm thermal}/\dot{E}_{\rm cool}$, and $\tau_{d}=(r^3/GM_{\rm enc})^{1/2}$.}. In m12, the halo is not fully in the ``hot mode'' given its relatively low virial temperature, so $\tau_{c}/\tau_{d}$ is steady at $\sim 10$ at large radii (actually highest in the core, where $\tau_{d}$ becomes very short). Here m13 is the ``most stable'' case (consistent with its lower sSFR), with $\tau_{c}/\tau_{d}\sim 100$ from $\sim 5-100\,$kpc. The higher density of halo gas in m14's core gives $\tau_{c}/\tau_{d}\sim 20$ within $\sim 50\,$kpc. Note these all exhibit rapid cooling, despite $\tau_{c}/\tau_{d}\sim 20-100$ being the lowest values in m13 \&\ m14, compared to the often-quoted critical value of $\sim 10$ in the literature \citep{2012MNRAS.420.3174S,2017ApJ...845...80V}. This partly owes to the structure being much more strongly multi-phase here -- the cooling gas has already cooled out of the thermally-unstable temperature range (to $T<10^{5}$\,K; not included in \fref{fig:coolingtime}), which makes the cooling rate of the remaining gas here appear longer.

Differences between physics variations are consistent with the SFR and cool gas mass plots above. Note the ``NoFB'' runs actually feature the longest cooling times in the center, as FB injects metals and dense gas into the hot phase, lowering its cooling time.

\vspace{-0.5cm}
\subsection{Cooling vs. Energy Input} \label{S:cool_t}

\begin{figure*}
\centering
\includegraphics[width=17cm]{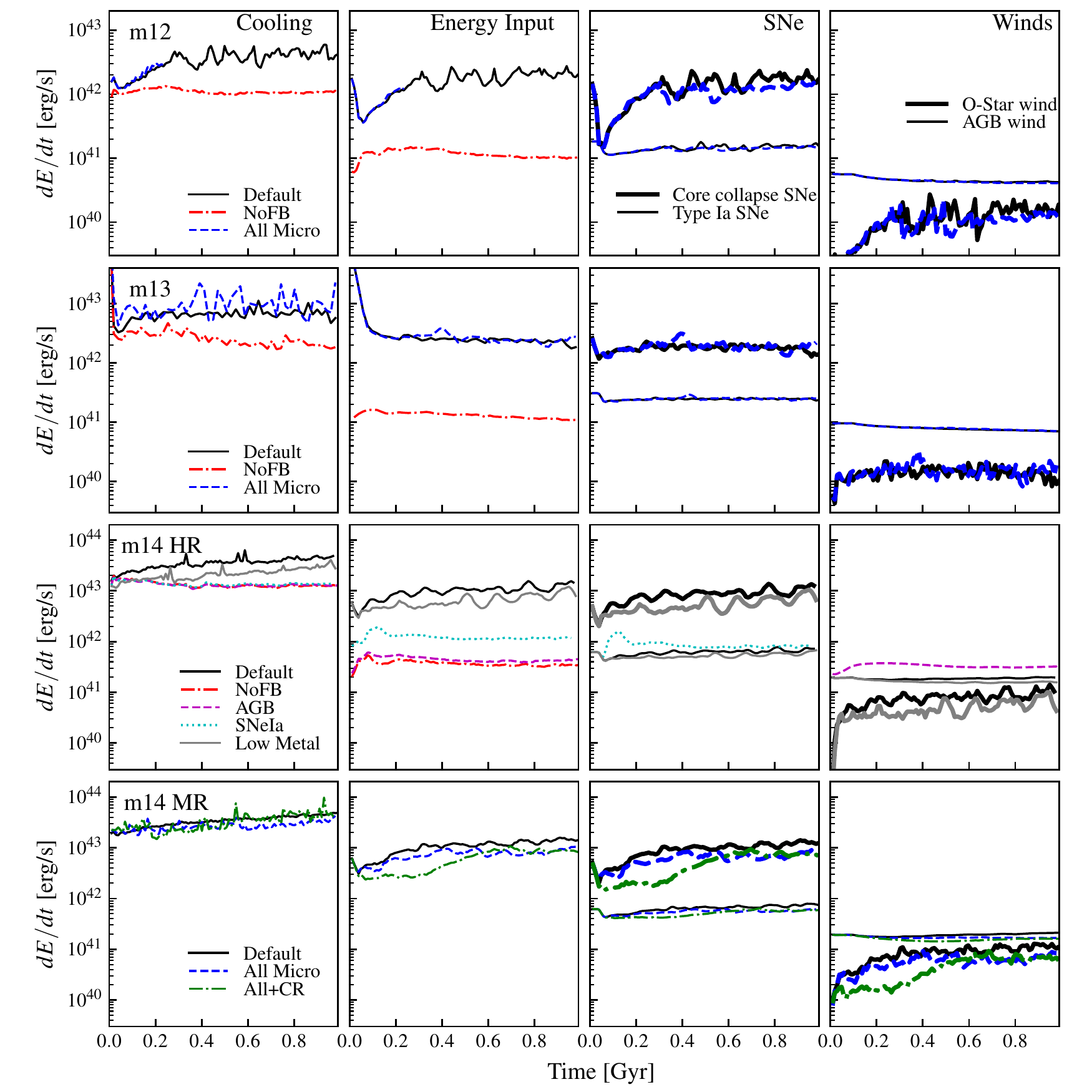}
\caption{Energy input, cooling, and feedback, as a function of time in m12 (top row), m13 (2nd row), m14 HR (3rd row), and m14 MR (bottom row) runs. {\bf Left column:} Total cooling rate within 30 kpc as a function of time. {\bf Middle left column:} Energy input rate, including UV background, cosmic ray heating, dust collisional heating, photo-electric heating, and the energy input from SNe and winds, within 30 kpc. {\bf Middle right column:} The energy input of core collapse SNe (thick lines) and type Ia SNe (thin lines) colored by simulation . {\bf Right column:} The energy input of O-Star winds (thick lines) and AGB winds (thin lines), colored by simulation. The plotted energy input from SNe assumes $10^{51}{\rm erg}$ per event, while winds energy input is calculated assuming the relative velocity between stars and gas is $300\,{\rm km\,s}^{-1}$ (upper bound). Cooling rates are always higher than heating rates. Core collapse SNe  input energy at  $\sim 1/3-1/4$ the cooling rate. Type 1a SNe input 20x  lower energy (time-averaged) compared to core-collapse in these runs, and AGB winds produce 2-3x lower energy injection compared to Ia's.  With suppressed metal-line cooling ("Low Metal"), the cooling rate is lowered by a factor of 2.}
\label{fig:coolingrate}
\end{figure*}

\fref{fig:coolingrate} compares cooling rates (all cooling channels added) and energy input rates (adding photo-ionization, photo-electric, cosmic ray, and the energy input from SNe and winds) within $<30\,$kpc. Cooling always exceeds heating. As expected, energy input in our ``Default'' run exceeds ``NoFB'' owing to higher energy input from e.g.\ SNe, but ``Default'' also maintains an even higher cooling rate. ``Low Metal'' has a factor $\sim 2$ lower cooling rate without metal-line cooling. Magnetic fields and viscosity produce negligible effects on their own. Conduction has weak effects here. We quantify the energy injection from each stellar feedback mechanism:\footnote{The plotted SNe energy input rate includes $10^{51} {\rm erg}$ per event. The plotted stellar wind energy input rate is actually an upper bound, since a relative velocity between gas and stars is assumed to be $300$\,km\,s$^{-1}$ for purposes of this post-processing estimate (it is calculated self-consistently in the code).} SNe Ia, O/B or AGB winds, and CRs contribute relatively little to total heating, with core-collapse SNe present (given that our galaxies are not quenched and have high SFRs). However even the core-collapse input is only $\sim 1/4-1/3$ of the cooling rate. SNe Ia input $\sim 20$x lower energy compared to core-collapse, and AGB winds produce a factor $\sim 2-3$x lower energy injection rate compared to Ia's.

The total X-ray luminosity of each run (in \fref{fig:xray}) scales with the cooling rates, as expected. In m12, the low halo temperature means the X-ray luminosity is significantly influenced by SNe heating; in m13 \&\ m14, the effects of stellar FB on the X-ray luminosity are small (most comes from the initial hot halo) \citep{2016MNRAS.463.4533V}.

\vspace{-0.5cm}
\subsection{Energetic Balance}
\begin{figure*}
\centering
\includegraphics[width=16cm]{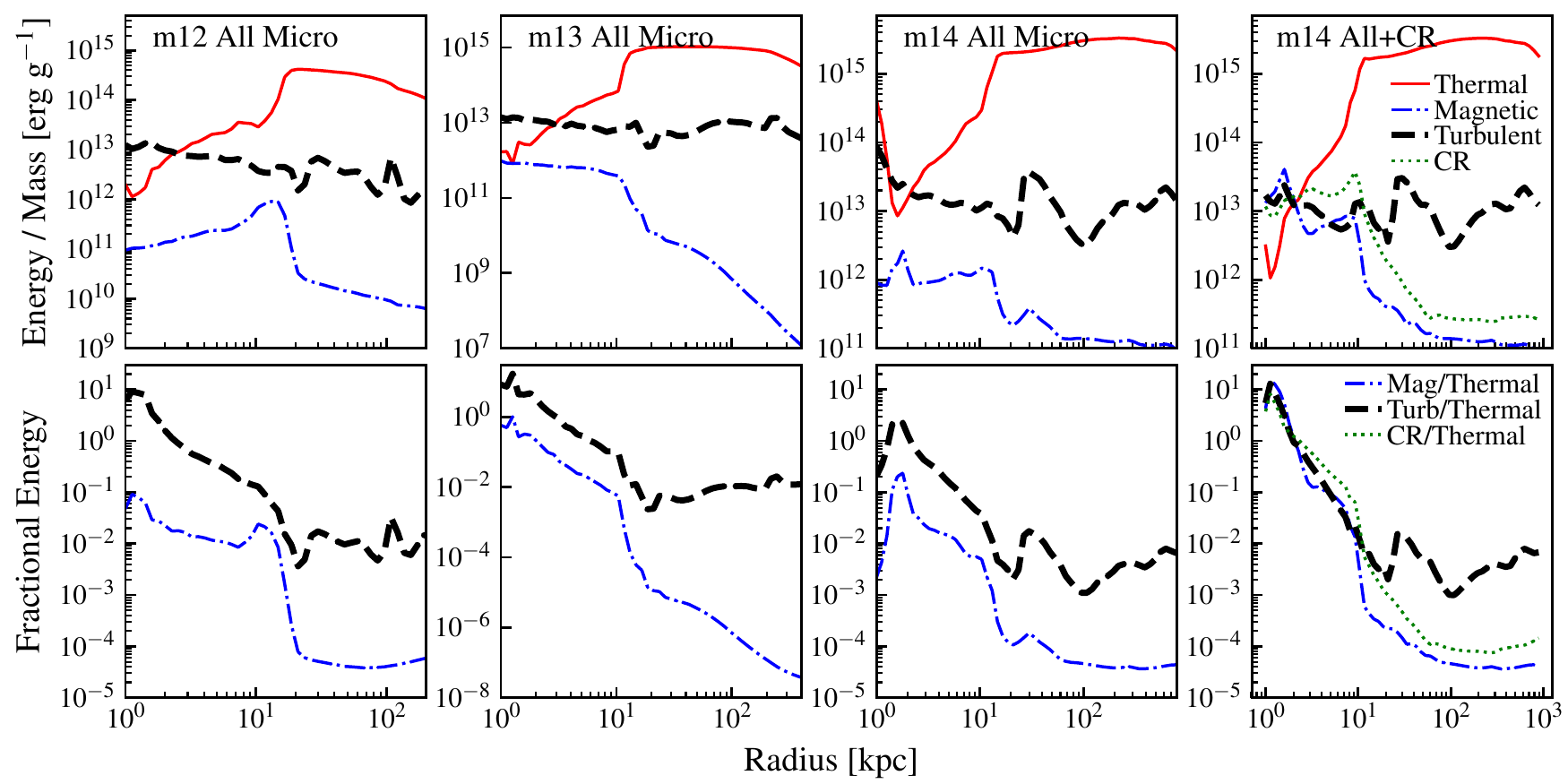}
\caption{{\bf Top row:} The comparisons of thermal, magnetic, CR and turbulent energy per unit mass, averaged over the 90-100th Myr. {\bf Bottom row:} The comparison of the ratio of magnetic, CR, and turbulent energy to thermal energy. For $r\lesssim 3$ kpc, magnetic energy is sub-dominant to turbulent energy, except in the `All+CR' run, where CR and magnetic energy reach equipartition with turbulent energy, showing that the turbulence is mostly super-Alfv\'enic and super-sonic.  At larger radius, $E_{\rm thermal}>E_{\rm turb}>E_{\rm CR}\gtrsim E_{\rm mag}$, so magnetic and CR energy have little effect on cooling flow. }
\label{fig:energy}
\end{figure*}

\fref{fig:energy} compares the  specific energy in thermal, magnetic, CR and turbulent forms averaged over 90-100th Myr.\footnote{Turbulent energies are measured using the method from \citet{2017MNRAS.471..144S} which attempts to separate turbulent motion from outflows and non-circular but bulk orbital motion.} Within a few kpc, turbulent energy dominates (the turbulence is super-sonic and super-Alfv\'enic), consistent with studies of the ISM inside galaxies \citep{2017MNRAS.471..144S,2018MNRAS.473L.111S}. At larger radii, thermal energy dominates (by at least $\sim 1\,$dex). Note at very large radii ($\gg 10\,$kpc), the magnetic energies are simply dominated by the ICs, since the flow is approximately laminar so the classical global dynamo amplification time is many orbital times ($\gg$\,Gyr). The same is true for the CR, as it does not have a chance to diffuse or stream to  $\gg 10\,$kpc within the simulation time.

\section{Discussion: Why Don't We Quench?} 
\label{s:discussion}

Here we briefly discuss why none of the mechanisms in \sref{S:physics} produces quenching, at any mass we survey.

\begin{figure}
\centering
\includegraphics[width=8.5cm]{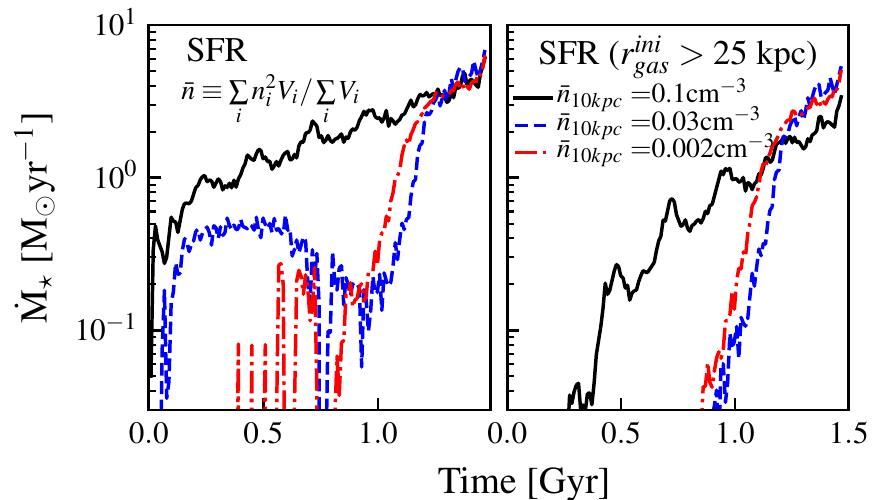}
\caption{The star formation rate of m12 runs with different core gas density within 10 kpc. $\bar{n}$ is defined as $\sum\limits_i n_i^2V_i/\sum\limits_i V_i$.  The SFRs are significantly suppressed if the core gas density is low. However, in all the cases, the SFR eventually catches up as the gas supplied by cooling flows builds up.}
\label{fig:agb_test}
\end{figure}

\subsection{Stellar Feedback}

\subsubsection{Young/Massive Stars}

Feedback from massive stars clearly reduces the rate at which gas {\em within the galaxy} turns into stars, self-regulating to a gas consumption time $\sim 1-2$\,dex longer than a case without feedback (bringing it into agreement with the observed Schmidt-Kennicutt  relation \citealt{1959ApJ...129..243S,1998ApJ...498..541K}), and drives local outflows from the disk \citep[e.g.][]{2013MNRAS.433.1970F,2017MNRAS.465.1682H,2018MNRAS.478.3653O}. However, cooling from the hot halo onto the galaxy eventually builds up the core gas mass and the SFR runs away.

Stellar feedback fails to suppress cooling in massive halos on long timescales for three reasons. 
{\bf (1)} Even with the elevated (much higher-than-observed) runaway SFRs, the total energy input from SNe is $\sim 1/4-1/3$ of the cooling rate (\fref{fig:coolingrate}). 
{\bf (2)} The energy is injected locally in the galaxy core, either as slow-moving (sub-$V_{\rm esc}$) cool gas or fast-but-tenuous hot gas, so is rapidly decelerated and does not couple outside the cooling radius (e.g.\ we verify that outflows in m14 rarely reach past $\sim 20-30\,$kpc
; see also \citealt{2015MNRAS.454.2691M,2017MNRAS.470.4698A} for more detail of the wind properties and the baryonic cycles in FIRE simulations). 
{\bf (3)} As is commonly seen in galactic fountains 
\citep[e.g.][]{2009IAUS..254..401S,10.1007/978-1-4419-7317-7_35,2011IAUS..277..273S,2013ApJ...764L..21F},  the outflows carry relatively dense, metal-rich gas into the halo, which {\em increases} the net cooling rate (\fref{fig:sfr} \&\ \fref{fig:coolingrate}) as it mixes with a larger mass of less-dense and lower-metallicity gas. The effect of metal enrichment on the cooling flows will be even clearer if the metal in the simulations were  solely from the stellar feedback, instead of partially from the initial conditions. In fact, we do see a factor of 2 lower SFR and core baryonic mass in the run with negligible initial metallicity (`Low Metal' run).

\fref{fig:agb_test} explicitly compares three variations of m12 that have initial central ($<10\,$kpc) gas fractions $f_{\rm gas}= \bar{n}_{\rm gas}/n_{\rm star} \approx 0.05,\,0.01,\,0.004$ (mean central densities $\bar{n}/{\rm cm^{-3}} = 0.1,\,0.03,\,0.002$). If there is little or no initial gas inside $<10\,$kpc, then (as expected) the initial SFR is suppressed strongly. More interestingly, we {\em also} see the SFR from gas with initial $r > 25\,$kpc is suppressed for $\approx 1\,$Gyr: this partially demonstrates how winds from SF in the disk (now absent) {\em enhance} cooling/inflow through enriching the halo gas. 
After $\approx 1\,$Gyr, however, cooling runs away and SF is dominated by the newly-cooled gas.

\subsubsection{SNe Ia \&\ AGB Winds}

Considering just Ia's or shock-heated AGB winds, we show in \fref{fig:coolingrate} that the feedback energy injection rate is even lower compared to cooling (by more than an order of magnitude), exacerbating problem {\bf (1)} above, and problems {\bf (2)} and {\bf (3)} remain. The explosions decelerate rapidly, and mixing the highly metal-rich gas\footnote{It is important to note that while AGB ejecta are much less metal-rich than pure Ia ejecta, it is still approximately solar (the mass-weighted mean stellar metallicity in massive galaxies) or somewhat more enriched in C and O (the primary coolants), and carries much larger mass, so it is less rapidly diluted. In fact, for a $\sim 10\,$Gyr old stellar population, the {\em total} metal return rate is {\em higher} by a factor of $\sim 4$ in AGB winds, compared to Ia SNe.} promotes cooling.

Some previous studies (see \sref{S:feedback}) appeared to reach different conclusions. However, these were largely based on simple analytic energetics arguments, so could not follow the non-linear effects of {\bf (2)} and {\bf (3)} above. Moreover, even for {\bf (1)}, although we find energy injection rates ``per star'' from SNe Ia and shocked AGB ejecta similar to these previous estimates, we find that clumping in the gas, mixing of winds, and cooling from larger radii enhances the central densities (and corresponding cooling rates) beyond the relatively low values assumed in those papers, rendering the heating insufficient.\footnote{Consider m12: Type Ia SNe and shocked AGB winds input energy at $\sim 1.5\times 10^{41} {\rm erg\, s}^{-1}$ and $\sim 5\times 10^{40} {\rm erg \,s}^{-1}$, respectively, roughly  consistent with the value quoted in \cite{2015ApJ...803...77C} for a similar-mass galaxy. However, in m12, the average effective gas density within $<10\,$kpc can be $\sim 1$\,dex higher than the $0.01\,{\rm cm^{-3}}$ assumed in \citet{2015ApJ...803...77C}. Note that clumping matters here: cooling rates scale $\propto n^{2}$, so the density that matters is $\bar{n} \equiv \langle n^{2} \rangle^{1/2}$, which is a factor of several higher in our runs than $\langle n \rangle$ inside $R_{\rm cool}$. Even assuming primordial gas (ignoring metal-cooling, {\bf (3)} above), and ignoring whether heating can reach large radii {\bf (2)}, the analytic scaling from \citet{2015ApJ...803...77C} then predicts $\dot{e}_{\rm cool}/\dot{e}_{\rm heat} \sim
40\,(f_{\rm gas}/0.01)\,(\bar{n}/0.1\,{\rm cm^{-3}})$ within the galaxy.} Even if the initial gas density within 10 kpc is lowered to $0.002\,{\rm cm}^{-2}$ (as shown in \fref{fig:agb_test}) , the stellar feedback (mostly SNe Ia given the SFR $\sim$ 0 within 1Gyr) can at most suppress the SFR for $\sim1$ Gyr, after which the core gas density builds up and the cooling runs away.

We note that the supernovae implementation deal with the unresolved Sedov-Taylor phases following \cite{hopkins:sne.methods}, which assumes negligible surrounding pressure \citep{1988ApJ...334..252C}. However, if a SN happens in CGM where the surrounding pressure is potentially high, it has to do extra pdV work as the blast wave expands, which  lowers the coupling momentum. Therefore, assuming negligible surrounding pressure only means overestimating the effect from SNe, but, even so, the effect of SNe is still limited.

\subsection{Magnetic fields, Conduction, \&\ Viscosity}

It is expected that magnetic fields alone cannot quench or suppress cooling flows: since they do not (directly) alter cooling, magnetic pressure would have to ``hold up'' the cooling gas in the halo, requiring un-physically strong fields (plasma $\beta \ll 1$, while $\beta \gg 1$ is observed and expected).\footnote{We confirm that we can, in principle, ``quench'' if we initialize enormously strong fields, but this requires magnetic field strengths exceeding self-gravity which simply ``explode'' the halo gas in a dynamical time.} Moreover even in that case, in 3D all field orientations are Rayleigh-Taylor unstable \citep[see e.g.][]{2007ApJ...671.1726S}. And extensive previous work has shown the fields have a small effect on galaxy-scale SF \citep[][and references therein]{2017MNRAS.471..144S}. Viscosity has equally small effect: it is weak and, if anything, slows and mixes outflows, slightly enhancing cooling.

Because of its strong temperature dependence (diffusivity $\propto T^{5/2}/\rho$), Spitzer-Braginskii conduction is expected to play a role only in the most massive halos, and we confirm this. The diffusion time for bulk heat transport across a distance $\sim R$ is $\sim 0.3\,{\rm Gyr}\,(R/10\,{\rm kpc})^{2}\,(n/0.01\,{\rm cm^{-3}})\,(T/10^{7}\,{\rm K})^{-5/2}$ -- only comparable to cooling times (inside the cooling radius $R_{\rm cool}$) in our most massive halo (m14). Moreover in a turbulent medium, eddies mix with approximate diffusivity $\sim v_{\rm eddy}(\lambda_{\rm eddy})\,\lambda_{\rm eddy}$; if we assume transsonic, Kolmogorov turbulence then micro-physical conduction dominates over turbulent only at scales $\lesssim 50\,{\rm pc}\,(\lambda_{\rm driving}/10\,{\rm kpc})^{1/4}\,(T/10^{6}\,{\rm K})^{3/2}\,(n/0.01\,{\rm cm^{-3}})^{-3/4}$. This is only larger than our resolution (and correspondingly, microphysical conduction dominates over both turbulent and numerical conduction with $\kappa\sim c_{s}\,\Delta x$ only) if $T \gtrsim 10^{7}\,{\rm K}\,(m_{g}/10^{4}\,{\rm M}_{\odot})^{0.2}\,(n/0.01\,{\rm cm^{-3}})^{0.3}$. So except in the outer regions of our most massive halo, the dominant heat transport/mixing is not Braginskii conduction. 

\begin{figure}
\centering
\includegraphics[width=8.5cm]{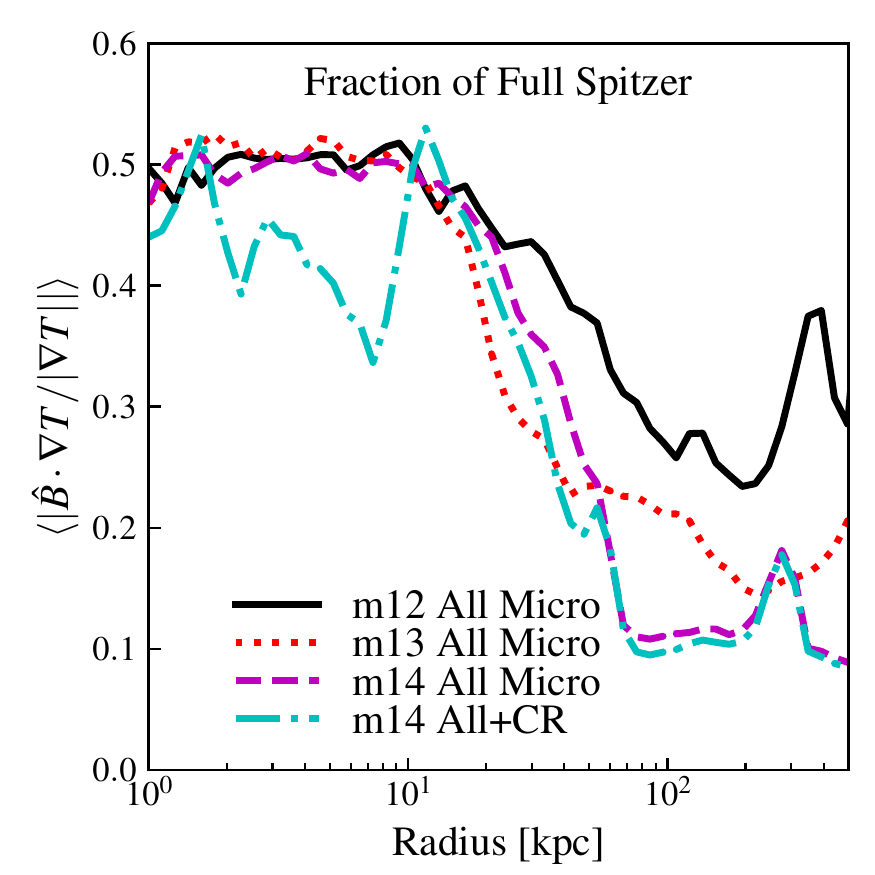}
\caption{The effective fraction of Spitzer conductivity, as a function of radius, estimated as the mass weighted average of $|\hat{B}\cdot\nabla T/|\nabla T||$. This fraction drops as a function of radius. In the core region, the conductivity is effectively half of the full Spitzer value, while it drops to 0.1-0.3 of it at large radii.}
\label{fig:compression}
\end{figure}

In m14, conduction is not negligible, but it only lowers the inflow rates and SFR from halo gas by a factor $\sim 2$. The effect is  modest because {\bf (1)} Lower temperatures and higher densities inside the core make conduction globally less efficient. {\bf (2)} Once something (e.g.\ dense outflows, turbulence) triggers thermal instability, cool ``clouds'' radiate efficiently and develop sharp density contrasts with the ambient medium so the conduction becomes saturated and can only out-compete cooling in the very smallest clouds \citep{1977ApJ...215..213M} -- in fact, recent work (\citealt{2016ApJ...822...31B,2017MNRAS.470..114A}; Hopkins et al., 2018, in prep) has shown that conduction often actually {\em increases} cold cloud lifetimes via cloud compression suppressing surface-mixing instabilities, in this limit. {\bf (3)} Magnetic fields modestly suppress perpendicular transport (quantified in \fref{fig:compression}, where in cores the conductivity is suppressed by a factor $\sim 2$, and in outskirts a factor $\sim 3-10$, levels (\citealt{2011ApJ...740...28V} and \citealt{2014MNRAS.439.2822W} argue will suppress many effects of conduction). 

Conduction may still be important to un-resolved small-scale thermal instabilities in hot halos (as noted above), but we note first that at finite resolution our numerical diffusion dominates (so if anything we over-estimate true conductivities), and second, most of the discussion in the literature on this question has focused on the regime where there is some {\em global} heat source injecting energy sufficient to offset cooling losses (where conduction plays the role of {\em local} heat transport into un-resolved clouds). When there is no heat input from feedback (or the heat input is less than the cooling rate, as in our default case here), then conduction does {\em not} significantly modify the consequences of the small-scale thermal instability (cold clouds ``absorb'' the conducted heat, condense and rain out efficiently; see \citealt{2015ApJ...799L...1V} ).

\subsection{Cosmic Rays from Stars \&\ LSS (not AGN)}

CRs from SNe have limited effect here: 
{\bf (1)} Direct heating (from e.g.\ streaming/hadronic interactions) cannot compete with cooling -- the total CR energy injection rate is $\sim 10\%$ of SNe, so order-of-magnitude less than cooling rates (\fref{fig:hotgas},\,\ref{fig:coolingrate}).
{\bf (2)} CR pressure is approximately in equipartition with magnetic energy (\fref{fig:energy}), so does not have a dramatic dynamical effect ``holding up'' the halo (though it can help accelerate some winds from massive stellar feedback at the disk).
{\bf (3)} Like other feedback from massive stars, CRs do help suppress the collapse/SFR from cold gas locally (see \fref{fig:sfr}, `All+CR' m14), but this leads to a pile-up of that gas (\fref{fig:hotgas}) from large-scale cooling, which runs away.
{\bf (4)} It requires some fine-tuning to make CRs do all their ``work'' around cooling-flow radii, rather than diffusing out.\footnote{While not shown here, we have experimented with e.g.\ variations in the CR diffusivity. If it is much lower ($\kappa_{\rm cr}\lesssim 10^{28}\,{\rm cm^{2}\,s^{-1}}$), CRs are trapped in the inner regions ($<10\,$kpc) and suppress SF from cold gas in the short-term more efficiently. But precisely because of this trapping and buildup of cold gas the CR energy is then mostly lost to catastrophic hadronic interactions, so the CRs have weaker long-term effects. For much higher diffusivity ($\kappa_{\rm cr}\gtrsim 10^{30}\,{\rm cm^{2}\,s^{-1}}$), CRs free-stream completely out of the halos with negligible interaction with gas.}

Although we do not directly model it, CRs from structure formation will have all these limitations as well. We find, for example, that we can ``quench'' via initializing an enormous CR density, but only if this is so large it overcomes gravity and blows out most of the halo gas. But this (a) does not resemble observed halos, and (b) is not possible from structure formation, since (by definition) only a fraction of the gravitational energy (in e.g.\ shocks) goes into CRs.

\subsection{Morphological quenching}
\label{s:morph_q}

Although galaxy colors and morphologies are correlated, that does not mean morphology {\em causes} quenching; we find that changing the stellar morphology of the galaxies here has very weak effects on their cooling/SF properties. 
{\bf (1)} Changing the morphology of stars has no direct effect on cooling rates, so even if it somehow quenched SF within the galaxy, cool gas would still pile up. For m14, this would give $>3\times10^{11}\,{\rm M}_{\odot}$ of molecular gas by the end of our simulation (\fref{fig:hotgas}), orders-of-magnitude higher than usually observed \citep[e.g.][]{2006A&A...454..437S,2015MNRAS.449..477P} (reference added). 
{\bf (2)} Moreover, if gas is self-shielding (i.e.\ reaches surface densities $\gtrsim 10\,{\rm M}_{\odot}\,{\rm pc^{-2}}$, requiring just $\sim 3\times10^{9}\,{\rm M}_{\odot}$ of gas within $<10\,$kpc, vastly less than that produced by the cooling flow), then it can cool to $T \ll 10^{4}$K, at which point Toomre $Q\ll 1$ will {\em always} be true in the cold gas for {\em any}  mass distribution with a physically-plausible rotational velocity. Indeed, the few known BCGs with large gas reservoirs $>10^{10}\,{\rm M}_{\odot}$, as predicted by our simulations here, all have observed SFRS $\sim 10-100\,{\rm M}_{\odot}\,{\rm yr^{-1}}$, like our m14, and obviously not quenched (see \citealt{2008ApJ...681.1035O}).
{\bf (3)} The effect of the stellar morphology on $Q$ is quite weak: somehow converting the {\em entire} MW stellar disk {\em and} all DM within $<10$\,kpc to a compact bulge or point mass would only increase the $Q$ at the solar circle by $\sim 50\%$ \citep[see e.g.][]{2011MNRAS.416.1191R}. In fact the stronger, but still weak, effect of changing an initial stellar disk to a bulge in \fref{fig:sfr} comes from slightly reducing the impact of stellar feedback on the inner halo (by making the stars older and less extended, so they have weaker feedback that reaches less far into the halo).
{\bf (4)} All our m13 and m14 runs, and of course observed massive galaxies, are completely bulge-dominated, yet still feature a ``cooling flow problem.''

\section{Conclusions} 
\label{s:conclusion}

In this paper we used high-resolution, idealized, isolated galaxy simulations including detailed physical treatments of star formation, stellar feedback, and ISM/CGM/ICM microphysics (cooling, magnetic fields, conduction, cosmic rays, etc.) to explore and quantify the quenching and cooling flow problems -- {\em in the absence of AGN feedback} -- across a range of halo masses from $\sim 10^{12}-10^{14}\,{\rm M}_{\odot}$. We specifically explored several ``non-AGN'' quenching or cooling-flow ``solution'' mechanisms, which have been previously proposed in the literature (e.g.\ feedback from old stellar populations in Type Ia SNe or shocked AGB ejecta, heat transport from the outer halo via conduction, cosmic rays from Ia's or structure formation, or ``morphological quenching''). None of these mechanisms resolve the fundamental problem of over-cooling and excessive star formation in massive galaxies, at any mass scale, that we simulate. The main effects of these physics are as follow:

\begin{itemize}
\item{Stellar feedback alters the balance of cold/warm gas and suppresses SFRs {\em for a given cold gas mass} (i.e.\ controls the location of galaxies on the Kennicutt-Schmidt relation). However it has only weak effects on cooling from the outer halo, and in fact tends to {\em enhance} cooling in the inner halo, as denser, more metal-rich ejecta mix and promote cooling in halo gas. This applies to all stellar feedback mechanisms (Ia's, AGB ejecta, feedback from young stars where present).}
\item{Magnetic fields and Braginskii viscosity have minor effects on the global cooling and inflow rates.}
\item{Conduction is, as expected, only important to bulk cooling/inflow in the most massive halos ($\ge 10^{14}\,{\rm M}_{\odot}$). Even there, the effects are modest, reducing inflow rates by a factor $\sim2$, owing to a combination of saturation effects, suppression by magnetic fields, and inefficient conduction in the cores once runaway cooling begins.}

\item{Cosmic rays from SNe (and shocked stellar winds) alone can, like other stellar feedback mechanisms, modestly reduce the SFR in cool gas already in/near the central galaxy, but their bulk energetics are insufficient to suppress cooling flows. We expect the same for CRs from structure formation.}
\item{Stellar morphology has essentially no effect on cooling rates and only weakly alters star formation either in pre-existing gas disks or in gas disks formed via runaway cooling. Making our galaxies entirely bulge-dominated does not make them quenched.}
\end{itemize}
 
Precisely because the effects of the above physics are weak, our summary of the quenching ``maintenance'' and/or ``cooling flow'' problems is consistent with many previous studies that treated some of the physics above in a more simplified manner: 
\begin{itemize}
\item{At all halo masses $\gtrsim 10^{12}\,{\rm M}_{\odot}$, we find efficient cooling of halo gas in cores, with cooling luminosities similar to observations (where available), but excessive cooling/cold gas masses and SFRs in central galaxies (from $\sim 5\,{\rm M}_{\odot}\,{\rm yr}^{-1}$ in $\sim 10^{12-13}\,{\rm M}_{\odot}$ halos, to $\sim 50\,{\rm M}_{\odot}\,{\rm yr}^{-1}$ by $\sim 10^{14}\,{\rm M}_{\odot}$).} 
\item{The excess gas comes, in an immediate sense, from an over-cooling core where higher densities and metallicities (enhanced by earlier generations of recycled galactic winds) produce rapid cooling and multi-phase CGM structure. Although the ``median'' cooling times in this core can be large compared to dynamical times ($t_{\rm cool}/t_{\rm dyn}\sim 100$), the fastest-cooling (denser, more metal-rich) material reaches $t_{\rm cool}/t_{\rm dyn} \ll 10-30$ -- this is what rapidly forms stars.}
\item{The core providing immediate fuel for SF can have a relatively small extent $\lesssim 30\,$kpc, but gas within this radius at later times originates from larger radii (up to $\sim 100$\,kpc) and migrates slowly inwards before ``runaway,'' so it may be possible to ``starve'' the cooling flow on longer timescales by suppressing just cooling/inflow from $\sim 30-100\,$kpc.}
\item{It is possible, in principle, to {\em temporarily} quench galaxies in this mass range if one can remove all their dense gas within $\lesssim 10\,$kpc. This will suppress star formation for $\approx 1\,$Gyr (surprisingly independent of halo mass), before runaway cooling from the extended halo restores the excessive SFRs.}
\end{itemize}
 
Our simulations have several limitations upon which future work could improve. We wished to construct idealized, controlled experiments so did not evolve fully-cosmological simulations -- we do not expect this to alter the fundamental conclusions above, but it could introduce additional important effects (e.g.\ stirring turbulence in halos via structure formation). We of course have finite resolution, so micro-physical phase structure in the CGM remains un-resolved and could alter the effective large-scale cooling rates. Our treatment of some physics (SNe, cosmic rays) is necessarily approximate (``sub-grid'') but it would require truly dramatic qualitative changes to our assumptions reverse our conclusions. Due to computational expense we could only run $\sim 20$ high-resolution simulations, which means we could not explore the potential diversity of properties of different halos at the same mass. 

Most obviously, we neglect AGN feedback, in various forms (jets, bubbles, winds, radiation, etc.). This work furthers the argument that something -- perhaps AGN -- beyond the ``known'' physics we include here, must be at work. We emphasize that many or all of the physics explored here (e.g.\ magnetic fields, cosmic rays, etc) may indeed play a critical role in AGN feedback, even if they do not dramatically alter cooling flows {\em absent} an AGN. In future work, we will explore generic classes of AGN feedback models proposed in the literature, in simulations incorporating the additional physics here which must (in reality) be present as well, in order to better understand the non-linear interactions of different feedback and ISM/CGM/ICM physics.


\acknowledgments
We thank Eliot Quataert for useful discussion.
The Flatiron Institute is supported by the Simons Foundation. Support for PFH was provided by an Alfred P.~Sloan Research Fellowship, NASA ATP Grant NNX14AH35G, and NSF Collaborative Research Grant \#1411920 and CAREER grant \#1455342. CAFG was supported by NSF through grants AST-1412836, AST-1517491, AST-1715216, and CAREER award AST-1652522, by NASA through grant NNX15AB22G, by CXO through grant TM7-18007, and by a Cottrell Scholar Award from the Research Corporation for Science Advancement.
DK was supported by NSF grant AST-1715101 and the Cottrell Scholar Award from the Research Corporation for Science Advancement. 
V.H.R. acknowledges support from UC-MEXUS and CONACyT through the postdoctoral fellowship.
Numerical calculations were run on the Caltech compute cluster ``Wheeler,'' allocations from XSEDE TG-AST130039 and PRAC NSF.1713353 supported by the NSF, and NASA HEC SMD-16-7592. \\

\bibliographystyle{mnras}
\bibliography{mybibs}

\appendix\label{appendix}
\normalsize
\section{Resolution study}\label{A:resolution}

Extensive resolution studies of the FIRE-2 feedback and physics models used in this paper are presented in \citet{2018MNRAS.480..800H}.
However since these did not address all of the specific questions in this paper, we performed a series resolution studies using our ``Default'' m14 halo, varying the mass resolution by a factor $\sim 300$ (\tref{tab:ic2}). The ``HR'' (``MR'') run in \tref{tab:ic2} matches the m14-HR (m14-MR) resolution in the text.

\begin{table}
\begin{center}
 \caption{Mass resolutions used in our studies for the default m14 Run}
 \label{tab:ic2}
 \begin{tabular}{@{\extracolsep{\fill}}ccccc}

 \hline
\hline
Resolution                     & $m_g$     &$m_h$    &$m_d$ &$m_b$  \\       
\hline 
 HR-HRS                  & 8e3-2e6    &4e7    & 8e3   &8e3\\
 HR                      & 8e3-2e6    &4e7    & 2.5e6 &2.5e6\\
 MR-MRS                  & 3e4-2e6    &4e7    & 3e4   &3e4\\
 MR                      & 3e4-2e6    &4e7    & 2.5e6 &2.5e6\\
 LR                      & 2e6        &4e7    & 1e7   &1e7\\

\hline
\hline
\end{tabular}
\end{center}
\begin{flushleft}

(1) Resolution name. LR: Low resolution. MR: Medium resolution. HR: High resolution. MRS(HRS):Medium (High) resolution initial stellar particles.
(2) $m_g$: Gas particle mass.
(3) $m_h$: Dark matter halo particle mass.
(4) $m_d$: Pre-existing stellar disc particle mass.
(5) $m_b$: Pre-existing bulge particle mass.
Note: All runs use the m14 halo with `Default' FIRE physics.
\end{flushleft}
\end{table}

\begin{figure}
\centering
\includegraphics[width=8.5cm]{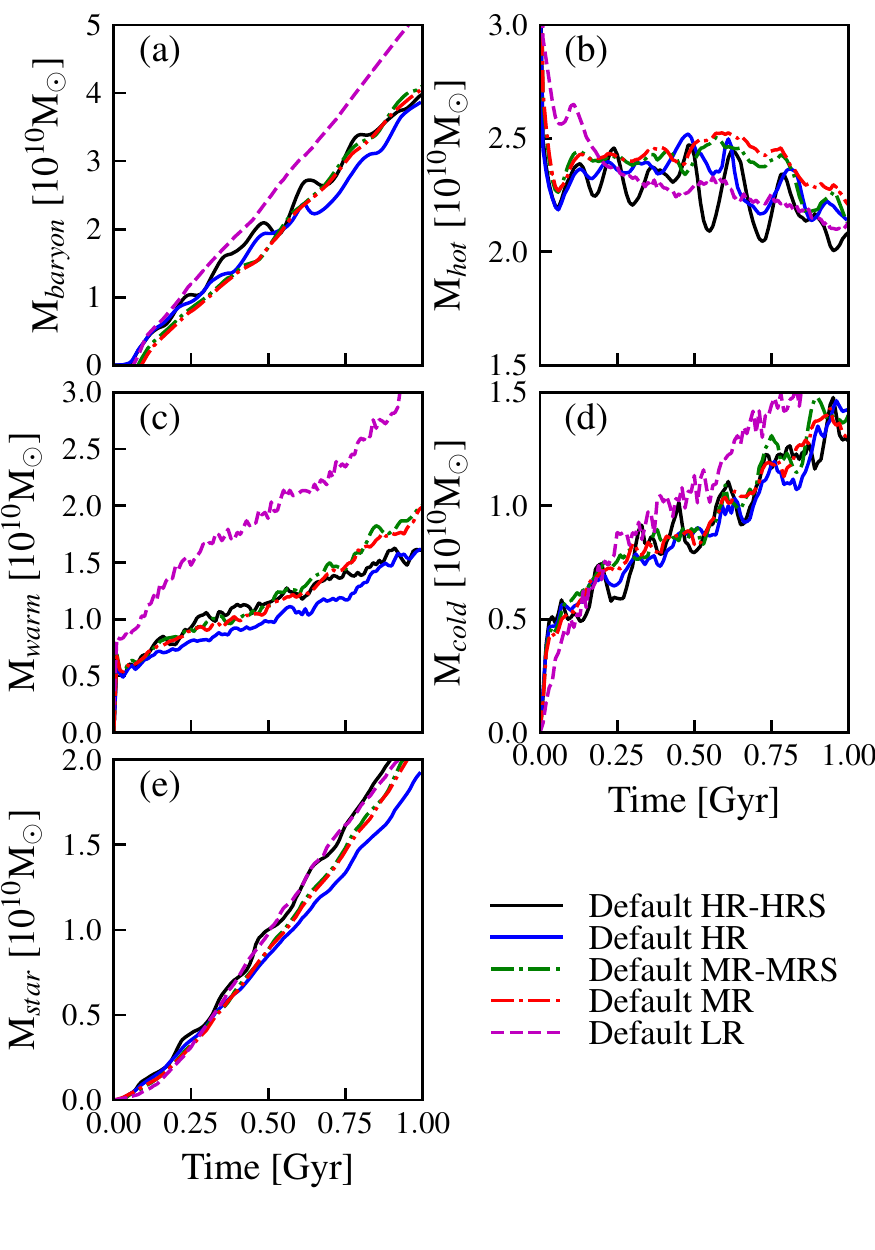}
\vspace{-0.7cm}
\caption{The comparison of (a) core ($<30$kpc) baryonic mass, (b) hot gas ($>10^6$K) mass, (c) warm gas ($8000-10^6$K) mass, (d) cold gas ($<8000$K) mass, and (e) stellar mass, for `Default' m14  runs at different resolutions. `MR' and `HR' runs behave very similarly. `LR' run, on the other hand has more gas accumulated in the warm phase.}
\label{fig:a_hot}
\end{figure}

\begin{figure}
\centering
\includegraphics[width=8.5cm]{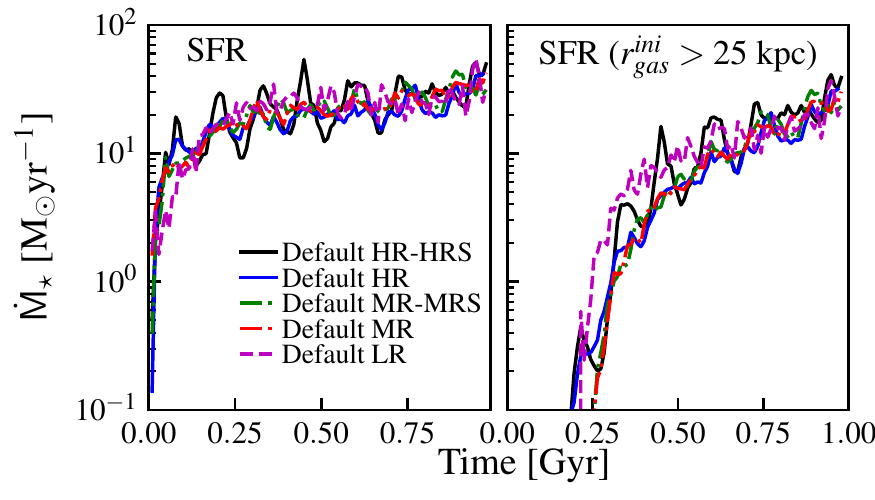}
\vspace{-0.4cm}
\caption{Comparison of total SFR, and SFR from gas initially outside 25kpc, for `Default' m14  runs with different resolutions. Runs at different resolutions have very similar SFRs.}
\label{fig:a_sfr}
\end{figure}

\begin{figure}
\centering
\includegraphics[width=8.5cm]{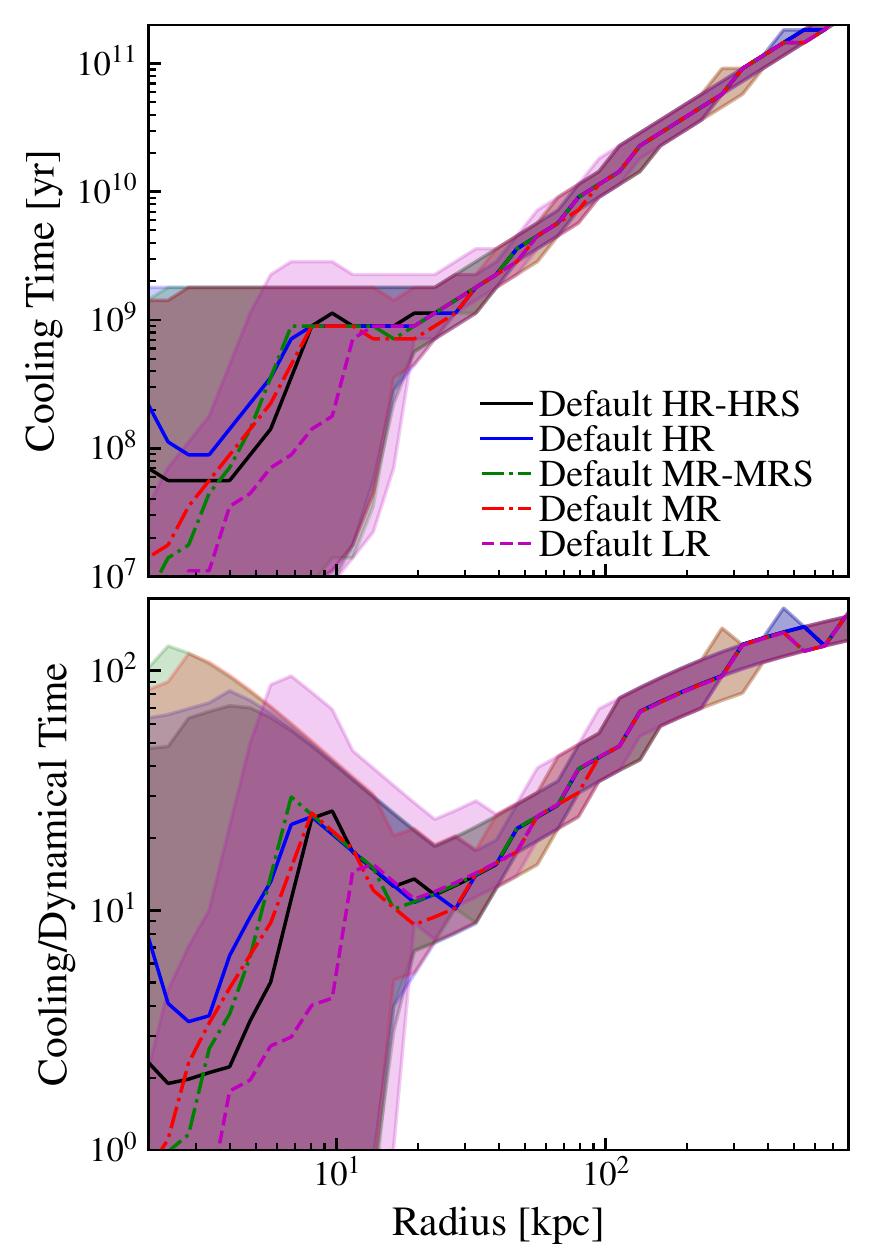}
\vspace{-0.4cm}
\caption{Cooling time, and cooling time over dynamical time, as a function of radius  for gas hotter than $10^5$K. `MR' and `HR' runs have very similar cooling properties. `LR' run, has slightly shorter cooling time at small radius.}
\label{fig:a_cool_r}
\end{figure}
\label{lastpage}

\begin{figure}
\centering
\includegraphics[width=8.5cm]{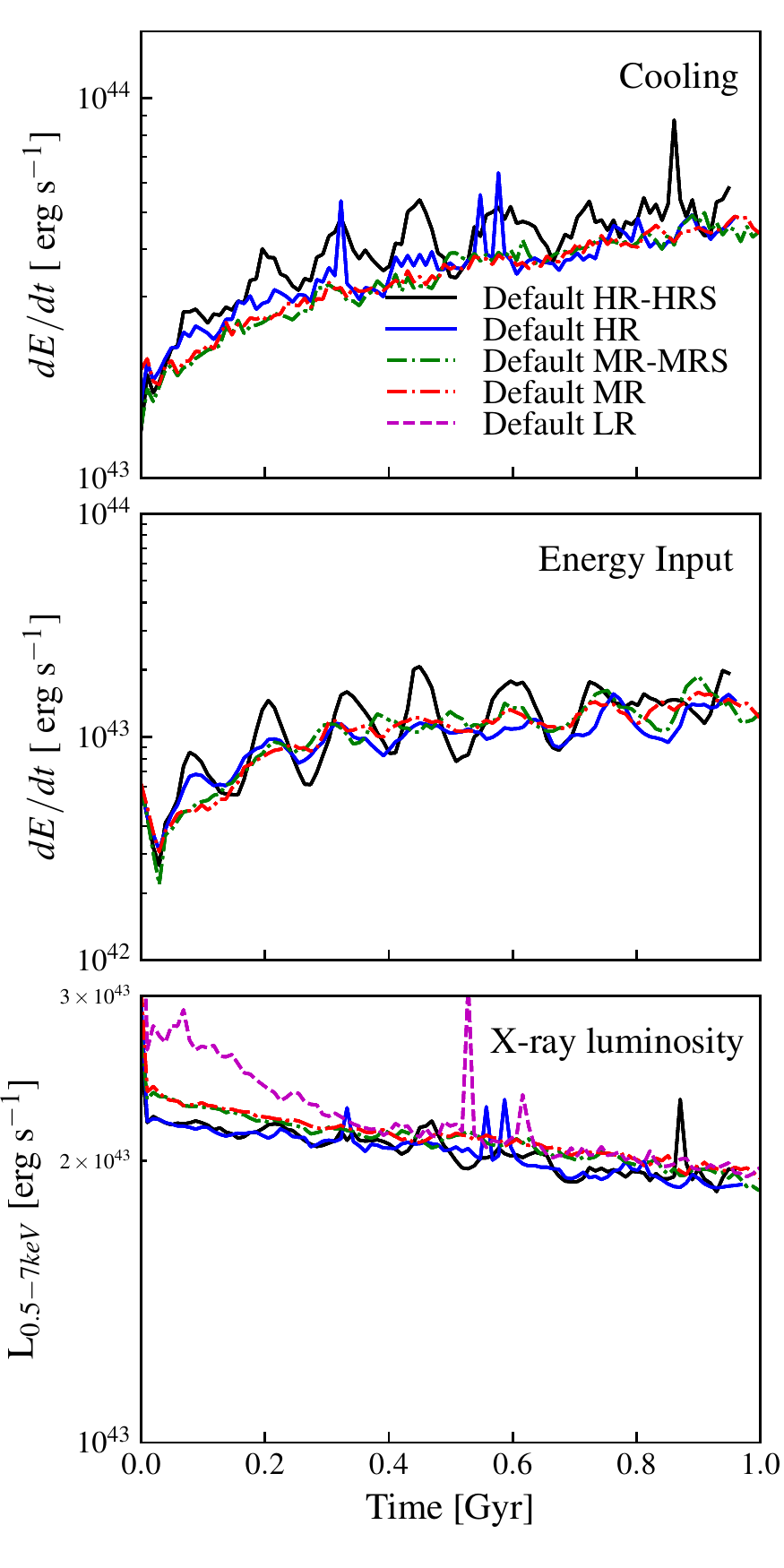}
\vspace{-0.4cm}
\caption{The upper row plots the cooling  and energy input rates within 30 kpc, as a function of time. The bottom row shows the  X-ray luminosity in the $0.5-7$ kev band.  Runs with different resolutions have very similar cooling and energy input rates.}
\label{fig:a_cool_t}
\end{figure}
\label{lastpage}

\fref{fig:a_hot} shows the evolution of the total core ($r<30$ kpc) baryonic, hot gas ($>10^6$K), warm gas ($8000-10^6$K), cold gas ($<8000$K), and stellar masses at different resolutions.
\fref{fig:a_sfr} shows the comparison of total SFR, and SFR from gas initially outside 25kpc. \fref{fig:a_cool_r} shows the comparison of cooling time and cooling time over dynamical time for gas within 30 kpc.
\fref{fig:a_cool_t} shows the evolution of cooling and energy input rates within 30 kpc, and the total X-ray luminosity of the whole halo. In these calculated properties, runs with resolution equal to or higher than that of our `MR' run do not differ significantly. The `LR' run on the other hand, exhibits more gas buildup in the warm phase at small radius, which leads to the shorter cooling time there. This, and the more detailed resolution studies referenced above, suggest the results are here are robust to resolution, at least over the dynamical range, that we explore here.

\end{document}